\documentclass[]{aa}  

\usepackage{graphicx}
\usepackage{txfonts}
\usepackage{indentfirst}
\usepackage{natbib}
\usepackage{textcomp}
\usepackage{ulem}
\bibpunct{(}{)}{;}{a}{}{,} 
\usepackage[scriptsize]{subfigure}
\usepackage{amsmath1}

\usepackage{color}
 
\begin{document}

  \title{Synthetic observations of first hydrostatic cores in collapsing low-mass dense cores.}
\subtitle{I. Spectral energy distributions and evolutionary sequence}  
   \author{B. Commer\c con
           \inst{1,2}
           ,
           R. Launhardt
 \inst{1}
  	  ,
 	  C. Dullemond
 \inst{3}
\and
Th. Henning
  \inst{1}
           }
   \offprints{B. Commer\c con}

\institute{Max Planck Institut f\H{u}r Astronomie, K\H{o}nigstuhl 17, 69117 Heidelberg, Germany\\
           \and
           Laboratoire de radioastronomie, UMR 8112 du CNRS, \'{E}cole normale sup\'{e}rieure et Observatoire de Paris, 24 rue Lhomond, 75231 Paris Cedex 05, France
   	              \email{benoit.commercon@lra.ens.fr}
\and	
             Zentrum fur Astronomie der Universit\H{a}t Heidelberg, Institut f\H{u}r Theoretische Astrophysik, Albert-Ueberle-Str. 2, 69120 Heidelberg, Germany\\
             }

   \date{Received december 22th 2011; accepted july 2nd 2012}

  \abstract  {The low-mass star formation evolutionary sequence is relatively well-defined  both from observations and theoretical considerations. The first hydrostatic core is the first protostellar equilibrium object that is formed during the star formation process.}   
  {Using state-of-the-art radiation-magneto-hydrodynamic 3D adaptive mesh refinement calculations, we aim to provide predictions for the dust continuum emission from first hydrostatic cores.}
  {We investigated the collapse and the fragmentation of magnetized 1 M$_\odot$ prestellar dense cores and the formation and evolution of first hydrostatic cores using the {\ttfamily{RAMSES }}\rm code. We used three different magnetization levels for the initial conditions, which cover a wide variety of early evolutionary morphology, e.g., the formation of a disk or a pseudo-disk, outflow launching, and fragmentation. We post-processed the dynamical calculations using the 3D radiative transfer code {\ttfamily{RADMC-3D}}. We computed spectral energy distributions and usual evolutionary stage indicators such as bolometric luminosity and temperature.}
      {We find that the first hydrostatic core lifetimes depend strongly on the initial magnetization level of the parent dense core. We derive, for the first time, spectral energy distribution evolutionary sequences from high-resolution radiation-magneto-hydrodynamic calculations. We show that under certain conditions, first hydrostatic cores can be identified from dust continuum emission at 24 $\mu$m and 70 $\mu$m. We also show that single spectral energy distributions cannot help in distinguishing between the formation scenarios of the first hydrostatic core, i.e., between the magnetized and non-magnetized models.}  
       {Spectral energy distributions are a first useful and direct way to target first hydrostatic core candidates but high-resolution interferometry is definitively needed to determine the evolutionary stage of the observed sources.}   


\titlerunning{}
\authorrunning{B. Commer\c con et al.}
   \maketitle


\section{Introduction}

It is well established that low-mass stars form from the collapse of  dense prestellar cores. The star formation process can be divided both observationally and theoretically into an evolutionary sequence. From observations, protostars are classified according to the parametrized properties of the observed spectral energy distribution (SED) and appearance. For instance, a  Class 0 source  \citep{Andre_et_al_1993}, which corresponds to the phase where the mass of the protostar is less than half the mass of the surrounding envelope, is defined according to its bolometric temperature $T_\mathrm{bol}$ \citep{Chen_et_al_1995} or to the ratio of bolometric to submillimeter luminosities $L_\mathrm{bol}/L_\mathrm{smm}$ \citep{Andre_et_al_1993}. At later stages, the Class I, Class II, and Class III sources can be classified according to the slope $\alpha_\mathrm{IR}=\mathrm{dlog}(\lambda F_\lambda)/\mathrm{dlog}(\lambda)$ between 2.2 $\mu$m and $10-25$ $\mu$m \citep{Lada_1987}. These classifications attempt to replicate the different evolutionary stages of a protostar. Observational ambiguities (e.g., projection effects, episodic accretion), however, imply that these classifications may not  always reflect the true evolutionary stage of an object \citep[e.g., ][]{Dunham_et_al_2010}.


The protostellar collapse phase is divided into two phases. First the prestellar dense core (gravitationally bound, no protostar, and no outflow) collapses isothermally and is able to freely radiate its gravitational energy into space, until it becomes dense ($\rho\sim 10^{-13}$ g cm $^{-3}$) and opaque enough for that radiation to be trapped. At this time, the gas begins to heat up and the first hydrostatic core (FHSC) is formed \citep{Larson_1969}.  The FHSC is optically thick and its temperature increases adiabatically until it reaches $\sim2000$ K, where H$_2$ dissociation starts. This endothermic reaction allows the gas to undergo a second collapse phase until stellar densities are reached, i.e., the formation of the second hydrostatic core (SHSC), the protostar that corresponds to the protostellar Class 0 evolutionary stage.   
The FHSC is thus a transition phase from prestellar cores to Class 0 objects and has not been yet identified conclusively because its observational appearance is unclear. The predicted timescales of the FHSC phase range from a few hundred years in spherical models \citep{Masunaga_Inutsuka_2000} to a few thousand years in very-low-mass non-magnetized rotating dense cores \citep{Tomida_2010b}.

In recent years, significant progress has been achieved in both theory and observation thanks to increasing computing power and new observational capabilities provided by {\it{Spitzer }}\rm and {\it{Herschel}}\rm. For instance, the large {\it{Spitzer }}\rm "Core to Disks" (c2d) Legacy Program survey \citep{Evans_et_al_2003} helped to increase the number of known protostars up to the point where first detections of FHSC can now be expected, given the short duration of the FHSC phase \citep{Enoch_et_al_2010,Dunham_et_al_2011}. {\it{Spitzer }}\rm also helped to identify a new class of low-luminosity objects: the Very Low Luminosity Objects \citep[VeLLOs, e.g., ][]{DiFrancesco_et_al_2007}. These are embedded in a dense envelope and have low intrinsic luminosity, $L<0.1$ L$_\odot$. It is still a matter of debate whether VeLLOs are dense cores undergoing first collapse or low-mass protostars with a low-accretion rate. In a recent study, \cite{Kim_et_al_2011} showed that at least in one case (the Bok globule CB130-1), the latter is the most likely explanation.
Prestellar dense core populations have been identified  over the past twenty years using space- and ground-based telescopes \citep[e.g., ][]{Ward-Thompson_et_al_1994,Motte_et_al_1998,Johnstone_et_al_2000,Nutter_2007}. 
More recently, {\it{Herschel }}\rm observations have identified hundreds of prestellar cores and Class 0 sources  \citep[e.g., ][]{Andre_et_al_2010,Bontemps_et_al_2010,Henning_et_al_2010}. 
 Combining different instrument data, \cite{Launhardt_et_al_2010} presented the results of a comprehensive infrared, submillimeter, and millimeter continuum emission study of isolated low-mass star-forming  cores, and identified nine Class 0 sources.
Finally, a few FHSC candidate detections have been reported in recent years thanks to the increasing spectral and spatial scale coverage \citep[][]{Belloche_et_al_2006, Chen_et_al_2010,Enoch_et_al_2010,Dunham_et_al_2011,Pineda_et_al_2011,Pezzuto_et_al}. Their confirmation remains controversial, however, because the expected observable signatures of FHSC are still uncertain.

In parallel,  enormous work has been put into developing numerical hydrodynamical models that include magnetic fields and radiative transfer \citep[e.g., ][]{Price_Bate_2009,Commercon_et_al_2010,Tomida_2010} but relatively little has been done in  producing synthetic observations that are directly comparable with actual observations. Most of the observables used to confront theory and observation remain statistical in nature, such as the initial mass function (IMF) or the binary distribution, whereas observations report mostly dust continuum and molecular line emission data. Therefore, it is vital that the interpretation of the data is sound, necessitating the use of non-symmetric models incorporating complex physics to derive predictions of dust continuum and molecular line emission observations. 

 For the early stages of star formation, \cite{Boss_Yorke_1995}, \cite{Masunaga_et_al_1998}, \cite{Omukai_2007}, \cite{Yamada_2009}, \cite{Tomida_2010b}, \cite{Tomisaka_2011}, and \cite{Saigo_Tomisaka_2011} made FHSC observational  predictions of either dust emission or molecular line emission. \cite{Young_Evans_2005} and \cite{Dunham_et_al_2010} also presented  evolutionary signatures of the formation of low-mass stars, but used a crude approximation of the FHSC and the evolution of the envelope during collapse. To date, there has been no systematic study of  observational predictions for FHSCs that explore  physical conditions, and in particular on the magnetization level. 
Magnetic fields have indeed been found to control the first collapse and fragmentation phases \citep[e.g., ][]{Hennebelle_Teyssier_2008,Commercon_et_al_2010} and strong magnetization seems to be a favored scenario for reproducing observational data of Class 0 multiplicity \citep{Maury_et_al_2010}. The ability of a disk to fragment depends ion the magnetization level, leading to potential changes in the SED.


In this paper, we derive SED evolutionary sequences and classical observational indicators such as $L_\mathrm{bol}$ and $T_\mathrm{bol}$ from state-of-the-art radiation-magneto-hydrodynamic (RMHD) 3D calculations of collapsing 1 M$_\odot$ dense cores with different initial magnetic field strengths. We address the questions of whether  or not SEDs can help in distinguishing physical conditions (e.g., magnetic field strengths) and how SEDs can help in targeting FHSC candidates. This paper is the first of a series and is associated with a companion paper in which we produce synthetic ALMA dust continuum emission maps (Commer\c con et al. {\it{in prep}}, hereafter Paper II).

The paper is organized as follows:  in Sect. 2,  we present the numerical codes we used and our post-processing methodology. The RMHD calculations are qualitatively presented in Sect. 3. In Sect. 4,  the RMHD calculations are post-processed to produce synthetic SEDs. In Sect 5., we discuss our results and potential observational tests.  Sect. 6 concludes our paper.

\section{Method}

\subsection{The {\ttfamily{RAMSES}} code}

For the dynamical RMHD calculations, we used the adaptive mesh refinement (AMR) code {\ttfamily{RAMSES}} \citep{Teyssier_2002}, which integrates the equations of ideal MHD \citep{Fromang_et_al_2006,Teyssier_et_al_2006} and the equations of radiation hydrodynamics under the gray flux-limited diffusion approximation \citep{Commercon_et_al_2011a}. The  {\ttfamily{RAMSES }}\rm code is well designed for collapse calculations  and has been extensively used in low-mass star formation studies \citep[e.g., ][]{Hennebelle_Fromang_2008, Commercon_et_al_2008, Hennebelle_Ciardi_2009, Commercon_et_al_2010}. It was also recently used in the high-mass star formation framework \citep{Hennebelle_et_al_2011,Commercon_et_al_2011c}.

\subsection{The {\ttfamily{RADMC-3D}} code}

To post-process the RMHD calculations, we used the 3D radiative transfer code  {\ttfamily{RADMC-3D}}\footnote{http://www.ita.uni-heidelberg.de/~dullemond/software/radmc-3d/}, developed by C. Dullemond. {\ttfamily{RADMC-3D}} can work with 3D AMR grids such as the  {\ttfamily{RAMSES}} grid. {\ttfamily{RADMC-3D}} has the capability to perform dust continuum calculations with anisotropic scattering and gas/molecular line transfer calculations (LTE or non-LTE). The dust temperature, emission, and absorption are computed using a thermal Monte Carlo method. {\ttfamily{RADMC-3D}} is able to compute images and spectra from dust continuum radiative transfer. For non-LTE line radiative transfer calculations,  {\ttfamily{RADMC-3D}} can estimate the level populations using approximate methods, such as the Large Velocity  Gradient (LVG) method \citep{Shetty_2011}. In this paper, we will only use the dust continuum features of  {\ttfamily{RADMC-3D}} and neglect scattering.

\begin{figure}[thb]
  \centering
  \includegraphics[scale=0.55]{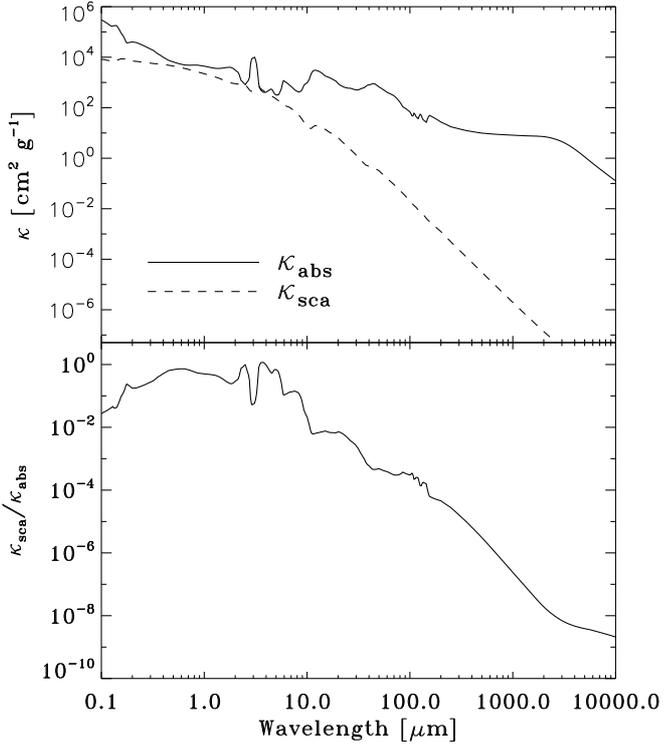}
\caption{{\it{Top:}} Scattering and absorption coefficient of \cite{Semenov_et_al_2003A&A} as a function of the wavelength for the  homogeneous
spheres model (with Fe/Fe+Mg=0.3 "normal" silicate composition). {\it{Bottom:}} Corresponding ratio between the scattering to absorption coefficients as a function of wavelength.}
\label{opacity}
\end{figure}

\subsection{Interface}

First, we assumed that  the dust and the gas are well coupled at high densities \citep[$\rho>10^{-19}$ g cm$^{-3}$ or $n>3\times10^{4}$ cm$^{-3}$; ][]{Galli_et_al_2002}  to estimate the dust temperature. Second, we assumed that the correct temperature is the 
gas temperature  computed in the dynamical RMHD calculations using the gray FLD approximation method. This second assumption is likely correct for FHSC formation and evolution, as shown recently by \cite{Vaytet_et_al_2011} for 1D spherically symmetric gray and multi-frequency  dense core collapse calculations.
To obtain more accurate temperatures, we could have used the thermal Monte-Carlo capabilities of {\ttfamily{RADMC-3D}}. During  the FHSC stage, however, the only sources  of photons are from the compressional work and the accretion luminosity, which are not trivial to extract from the RMHD calculations. 

Thanks to the AMR grid capabilities of  {\ttfamily{RADMC-3D}}, we directly loaded the AMR structure of  {\ttfamily{RAMSES}} within  {\ttfamily{RADMC-3D}}. We thus did not lose any information between the RMHD and the post-processing radiative transfer calculations. We show in Appendix \ref{appendixa} the importance of using the AMR grid instead of remapping the data onto a uniform grid. 

We  used the low-temperature
opacities of \cite{Semenov_et_al_2003A&A}  for the homogeneous
sphere model (with Fe/Fe+Mg=0.3 "normal" silicate composition). For the RMHD calculations, we used a parametrization of the gray Rosseland and Planck means as functions of the gas temperature and density. For the post-processing radiative transfer calculations, we used the corresponding frequency-dependent opacities. Figure \ref{opacity} shows the absorption and scattering opacities we used, as well as the ratio of scattering to absorption coefficients. The scattering coefficient is always less than the absorption coefficient, while their ratio is greater than 10\% for visible to long near-infrared wavelengths (i.e., $0.2$ $\mu$m $<\lambda<8$ $\mu$m). Since we neglected scattering, we may have overestimated the amount of radiation that escapes at shorter wavelengths. Neglecting scattering remains a good approximation, however, since in this study we focus on the internal luminosity of collapsing dense cores at a stage earlier than the protostar+disk stage where SEDs peak at shorter wavelengths  \citep[e.g., ][]{Dunham_et_al_2010}

\begin{figure*}[ht]
  \centering
  \includegraphics[scale=0.65]{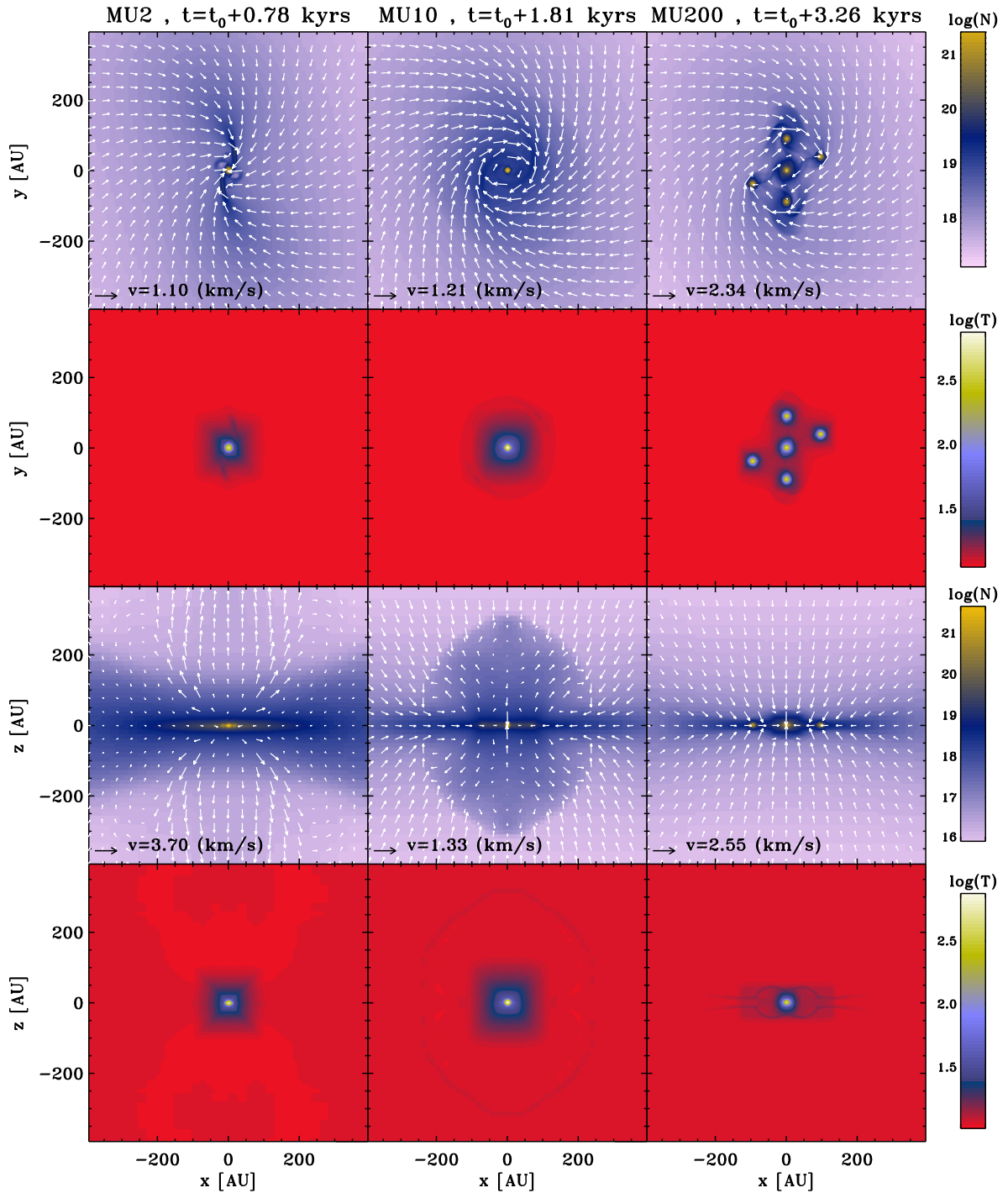}
\caption{Column density and temperature maps in the $xy$-plane (two upper rows) and $xz$-plane (two lower rows) for the three RMHD calculations: MU2 (left), MU10 (middle), and MU200 (right). Scales are logarithmic, and column density $N$ is given in cm$^{-2}$.}
\label{RMHD_res}
\end{figure*}

\section{Radiation-magneto-hydrodynamics models}
\subsection{Initial conditions}

We used similar initial conditions as in \cite{Commercon_et_al_2010}. We considered 1 M$_\odot$ uniform density  and temperature sphere in solid body rotation about   the  $z$-axis. The initial radius of the sphere is $R_0=4.93\times10^{16}$ cm ($\sim 3300$ AU). The initial angular velocity is set using a ratio of the rotational to gravitational energies $\beta=0.045$. To favor fragmentation, we introduced  an $m=2$ azimuthal density perturbation with an amplitude of $10\%$. 
The corresponding free-fall time and orbital time are $t_\mathrm{ff} \sim 33$ kyr and  $t_\mathrm{orb} \sim 1.85 \times10^2$ kyr, respectively. The initial density is $\rho_0=3.97\times 10 ^{-18}$ g cm$^ {-3}$ and the initial temperature is $T_0=11$ K. The ratio of specific heats $\gamma$ was set to $5/3$. In this study, we did not include any additional microphysics such as H$_2$ dissociation. Therefore, our models are only able to follow the dynamics up to a temperature of 2000 K, at which the second collapse starts. 

The magnetic field is initially uniform and parallel to the rotation axis. The strength of the magnetic field is expressed in terms of the mass-to-flux to critical mass-to-flux ratio
$\mu=(M_0/\Phi)/(M_0/\Phi)_\mathrm{c}$.
In this study, we explore three magnetization degrees: $\mu=200$, which corresponds to a quasi-hydrodynamical model (model MU200), $\mu=10$ for an intermediate magnetic field strength (model MU10), and $\mu=2$ for a strong magnetic field case (model MU2).

We used a coarse grid of $64^3$ cells with ten levels of refinement, which corresponds to a maximum effective resolution of 0.2 AU. Calculations were performed using
the HLLD Riemann solver \citep{Miyoshi_Kusano_05}. Following up on former studies \citep{Commercon_et_al_2008,Commercon_et_al_2010}, we imposed at least 12 cells per Jeans length as a grid refinement criterion.

\begin{figure*}[t]
  \centering
  \includegraphics[scale=0.78]{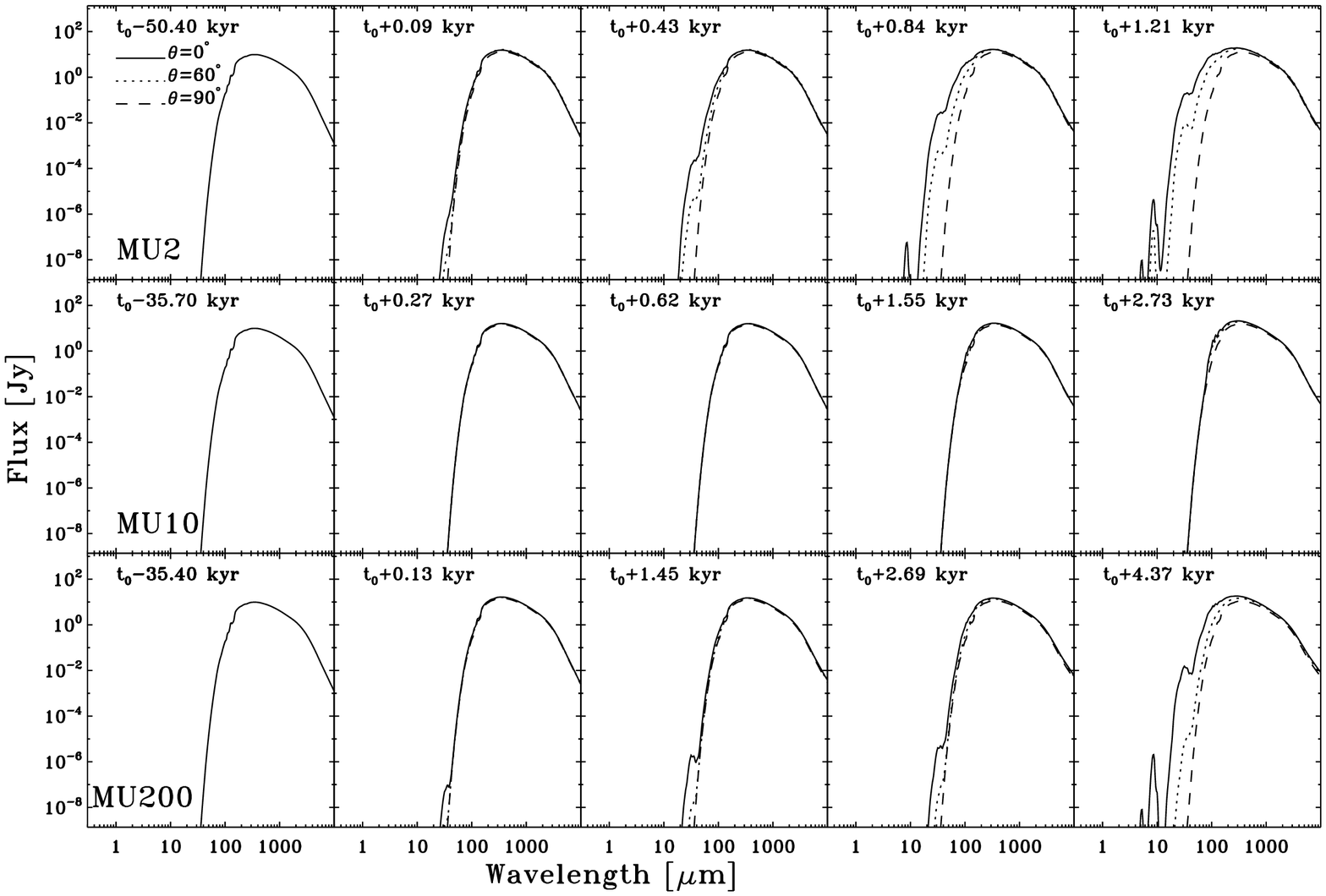}
\caption{Spectral energy distributions for the three models at different times: MU2 (top row), MU10 (middle row), and MU200 (bottom row). The first panels in the left column show the SEDs obtained from the initial conditions. SEDs are computed at each epoch for three different inclination angles: $\theta=0^{\circ}$ (pole-on view, solid line), $\theta=60^{\circ}$ (dotted line), and $\theta=90^{\circ}$ (edge-on view, dashed line).}
\label{sed}
\end{figure*}


\subsection{Qualitative description\label{t0}}

In the following, we describe only qualitatively the results of the three RMHD calculations since the focus of this paper is  the observational characteristics of the models and not  their physical quantitative description. We synchronized the calculations at time $t_0$ at which the maximum level of refinement is reached, which corresponds roughly to the formation of the FHSC. Column density and temperature maps for the three calculations are shown in Fig. \ref{RMHD_res}. All fragments that are formed in the calculations are FHSCs.

\subsubsection{The strongly magnetized case - $\mu=2$}

In the strongly magnetized case, fragmentation is totally suppressed and  no rotationally supported disk is formed because of the efficient magnetic braking \citep[e.g., ][]{Joos_et_al_2012}. At time $t_0+0.78$ kyr (with $t_0 =50.4$ kyr), there are $1.66 \times 10^6$ cells in the whole computational domain. An outflow is launched along the rotational axis with velocity $\sim 3.7$ km s$^{-1}$ and it extends up to $\sim 550$ AU (not seen in the zoom in Fig. \ref{RMHD_res}).  Only one  object is formed, surrounded by a thick pseudo-disk region. The mass of the FHSC is $\sim$ 0.06 M$_\odot$ and its mean temperature is $\sim375 $ K ($T_\mathrm{max} \sim 630$ K). The heated region around the central FHSC is almost spheroidal, which cannot be obtained in models using a barotropic approximation. In the latter case, the warm region would correspond to the dense region (the temperature is set by the density), i.e., the pseudo-disk \citep[see][]{Commercon_et_al_2010}. These differences in the temperature distribution are fundamental to the SED estimates. 
 We ran the MU2 calculations until time $t_0+1.2$ kyr, where $T_\mathrm{max} \sim 1160$ K and the outflow extends to $\sim 850$ AU, i.e., just before the start of the second collapse. At temperature $T>1200$ K, dust begins to evaporate, which considerably lowers the opacity by more than two orders of magnitude \citep[e.g., ][]{Lenzuni_et_al_1995}. The high temperature region is then able to radiate efficiently and the temperature increases fast towards 2000 K, at which time H$_2$ dissociation starts. 

\subsubsection{The intermediately magnetized case - $\mu=10$}

In the intermediately magnetic field strength model, there is no fragmentation, but an outflow of extent $\sim 300$ AU and velocity $\sim 1$ km s$^{-1}$ is launched and a rotationally supported disk of radius $\sim 100$ AU is formed. This result agrees well with previous studies of \cite{Hennebelle_Ciardi_2009} and \cite{Commercon_et_al_2010}. At time $t_0+1.81$ kyr (with $t_0 = 35.7$ kyr), there are $1.3 \times 10^6$ cells in the whole computational domain. 
The mass of the FHSC is $\sim$ 0.06 M$_\odot$ and its mean temperature is $\sim490 $ K ($T_\mathrm{max} \sim 810$ K).  From the temperature maps, we distinguish two shocked regions (i.e., where there is a weak spike in gas temperature) in addition to the accretion shock on the central FHSC itself. In the $xy$-plane, the shocked region  corresponds to the border of the disk. In this optically thin region, it is then easy to see the spike in gas temperature that is characteristic of radiative shocks and has a length of a few photon mean-free paths \citep[e.g., ][]{Mihalas_book}. In the $xz$-plane, we identify a shock at the outflow border, with a spike in the gas temperature of the same amplitude as at the disk border. In comparison with the MU2 model, the outflow is broader and denser, and the vertical extent of the pseudo-disk is strongly diminished.
The calculations of the MU10 model were run until $T_\mathrm{max}\sim1200$ K at time $t_0+2.94$ kyr, just before the second collapse is expected to begin.

\subsubsection{The very weakly magnetized case -  $\mu=200$}

In  the very weakly magnetized model, the collapsing core forms a prestellar disk that fragments into several objects (five at time t$_0 +3.26$ kyr), but no outflow is launched because of the weak initial magnetic field. The disk radius extends up to $\sim 150$ AU. At time $t_0+3.26$ kyr (with $t_0 = 35.4$ kyr), there are $1.95 \times 10^6$ cells in the whole computational domain. The mass of the fragments is about the same: $M\sim 0.043$ M$_\odot$ for the central fragment and $M\sim0.036-0.039$ M$_\odot$ for the orbital fragments. The mean temperature is $T\sim270$ K for the central fragment and $T\sim245-263$ K for the orbital fragments. Heating of the gas around the central fragment is much less efficient than in the two previous simulations.
In addition, we see from the temperature maps in the $xz$-plane that the radiative shock at the vertical borders of the disk is sharper and is located closer to the center compared to the MU10 run. The infall velocity and thus the shock strength are also higher. Before being accreted by a fragment, gas has already lost a part of its gravitational energy through that shock, and the accretion shock on the fragment is then weaker. The MU200 calculations were halted before the onset of the second collapse, the first core lifetime being relatively long for the hydrodynamical case \citep{Tomida_2010b}.
\\

From this qualitative description, we clearly see that the three models are different and relevant to the various conditions that may be found at the FHSC stage. In the following, we will see that these differences are also expressed in synthetic SEDs of these collapsing objects.

\section{Synthetic observations}

\subsection{Definitions of classical physical stage indicators}

The  SEDs of the models were used to calculate classical observational signatures. We monitored for each model the temporal evolution of the bolometric luminosity $L_\mathrm{bol}$, the bolometric temperature $T_\mathrm{bol}$, and the ratio of bolometric to submillimeter luminosity $L_\mathrm{bol}/L_\mathrm{smm}$. The bolometric luminosity $L_\mathrm{bol}$ is defined as
\begin{equation}
L_\mathrm{bol} = \int_0^{\infty} 4\pi d^2S_\nu d\nu,
\end{equation}
where $d$ is the distance to the object, $S_\nu$ is the flux density, and $\nu$ is the frequency. The bolometric temperature $T_\mathrm{bol}$ is defined following \cite{Myers_Ladd_1993}, i.e., as the temperature of a blackbody with the same mean frequency as the observed SED, which can be approximated by
\begin{equation}
T_\mathrm{bol} = 1.25\times10^{-11}\frac{\int_0^{\infty} \nu S_\nu d\nu}{\int_0^{\infty} S_\nu d\nu} \hspace{3pt}\mathrm{K}.
\end{equation}
Following \cite{Andre_et_al_1993}, we also calculated the ratio $L_\mathrm{bol}/L_\mathrm{smm}$, with the submillimeter luminosity estimated as 
\begin{equation}
L_\mathrm{smm} = \int_0^{\nu=c/350\mu\mathrm{m}} 4\pi d^2S_\nu d\nu.
\end{equation}
These three quantities are commonly used to indicate the physical stage of observed sources. $T_\mathrm{bol}$ and $L_\mathrm{bol}/L_\mathrm{smm}$ are both used to define class boundaries. For our purpose, only the boundary between Class 0 and Class I matters. \cite{Andre_et_al_1993} defined Class 0 sources as objects with  $L_\mathrm{bol}/L_\mathrm{smm}\leq 200$, while \cite{Chen_et_al_1995} defined Class 0 objects with $T_\mathrm{bol}<70$ K. Both classifications are problematic for models because one has to define the region over which to integrate the luminosity. Finally, VeLLOs are defined as objects with an internal luminosity $L_\mathrm{int}<0.1$ L$_\odot$, where $L_\mathrm{int}$ is  the luminosity of the object in excess of that supplied by the interstellar radiation field \citep{DiFrancesco_et_al_2007}.

\begin{figure}[t]
  \centering
  \includegraphics[scale=0.6]{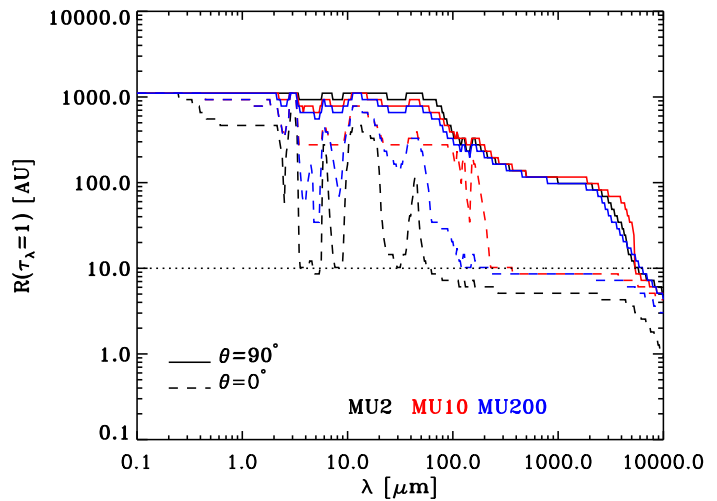}
\caption{Radius at which the optical depth $\tau_\lambda=1$ as a function of the wavelength for the three models at the same time as in Fig. \ref{RMHD_res}: MU2 (black), MU10 (red), and MU200 (blue). The solid line shows optical depth with an inclination of $\theta=90^{\circ}$, and the dashed line that of $\theta=0^{\circ}$. The dotted line at $R=10$ AU roughly represents the FHSC's outer boundary.}
\label{Rtau}
\end{figure}

\subsection{Spectral energy distribution time evolutionary sequence}
 
 In this subsection, we qualitatively describe the SEDs obtained for the three models. The fluxes of each SED shown below were computed and integrated over a lobe size of  3000 AU x 3000 AU (about half the initial size of the core) and observed at a distance of 150 pc (a typical distance to star-forming regions like Taurus).  These choices give an aperture of $\sim 20$'' for our synthetic observations, which compares well with current instrument resolutions. Our choice of fixing the same object size  for all wavelengths is justified because observations often report flux over a given area whatever the beam size. In addition, there are many different instruments that report star-forming region observations  and it is impossible to match all beam sizes in a single SED.

Figures \ref{sed} shows the SEDs at different epochs for the three models. For each epoch, SEDs are presented for three different inclination angles: $\theta=0^{\circ}$ (pole-on view, solid line), $\theta=60^{\circ}$ (dotted line), and $\theta=90^{\circ}$ (edge-on view, dashed line). We also computed SEDs for models at an inclination angle of $\theta=45^{\circ}$, but these were indistinguishable  from those obtained with an inclination of $\theta=0^{\circ}$.
 As long as $t<t_0$ in all models (see sec. \ref{t0} for the definition of $t_0$), SEDs are indistinguishable from the initial one and no emission is found at wavelengths shorter than 30 $\mu$m. This behavior is not surprising because the FHSC has not yet formed and the gas is still isothermal at $T=11$ K. 
 
 Surprisingly,  the SEDs of the MU2 and MU200 models have similar time evolutionary sequences compared to the MU10 model, despite their totally different physical conditions. In the MU2 and MU200 models, the departure from the initial conditions starts when $t>t_0$ and the flux increases at all wavelengths as temperature increases within the FHSC. The peak of the SED, however, does not change after the FHSC formation. We also find that a significant amount of radiation is emitted between 20 $\mu$m and 100 $\mu$m in the late evolution of the MU2 and MU200 models, in contradiction to previous studies \citep[e.g., ][]{Masunaga_et_al_1998,Omukai_2007}, which found no observable emission below $30 - 50$  $\mu$m. This difference can be explained by geometrical effects. \cite{Masunaga_et_al_1998} and \cite{Omukai_2007} assumed spherical symmetry, whereas we considered a full 3D structure. In our models, the envelope seen with inclination angle $\theta<60^{\circ}$ is much less dense than that of spherical models. In the same manner, the flux we obtain with $\theta=90^{\circ}$ is strongly reduced in this wavelength range because the envelope has reprocessed the radiation emitted by the FHSC.  
 Interestingly, the MU10 model does not show any evolutionary sequence even at the latest stage before the second collapse, independent of the inclination angle. In the edge-on view ($\theta=90^{\circ}$), this lack of variation is not surprising, since the MU10 model combines a disk and a pseudo-disk structure. The radiation emanating from the FHSC is thus fully  reprocessed. But even looking pole-on through the outflow, no flux below 30 $\mu$m is seen. From Fig. \ref{RMHD_res}, we can see that the outflow in the MU10 model is much denser than that in the MU2  model; it acts as an envelope in which radiation from the FHSC is reprocessed. 

Figure \ref{Rtau} shows the radius at which the optical depth $\tau_\lambda=1$ (the transition from the optically thin to thick regime) as a function of the wavelength for the three  models at the same time as in Fig. \ref{RMHD_res} and for two inclinations: $\theta=90^{\circ}$ (solid line, edge-on) and $\theta=0^{\circ}$ (dashed line, pole-on). The dotted line at $R=10$ AU roughly represents the FHSC outer boundary. The sharp jumps observed in the wavelength range between $\sim2$ $\mu$m and $\sim50$ $\mu$m are due to jumps in the opacity itself owing to ice or silicate for instance (see Fig. \ref{opacity}). Because the opacity increases when the wavelength decreases, the optically thick radius is larger at short wavelengths. For $\theta=90^{\circ}$, the optical depth is about the same in the three models, since the pseudo-disk or the disk play the same role in obscuring radiation from the FHSCs. The central FHSC boundary can thus only be seen in an edge-on view at  wavelengths $> 5$ mm. The disk or pseudo-disk features alone (typical size of  about 100 AU) are  nevertheless observable at shorter millimeter wavelengths, although these features are not reliable diagnostics of the presence of  FHSCs, since they will not disappear as the second collapse proceeds. For $\theta=0^{\circ}$, the radius at which the optical depth reaches unity corresponds to the FHSC radius for wavelengths $> 100 $ $\mu$m, and even marginally for shorter wavelengths 3 $\mu$m$<\lambda<$ 100 $\mu$m for the MU2 model.  At inclination angle $\theta<60^{\circ}$ (since the $\theta=0^{\circ}$ SED closely resembles the $\theta=60^{\circ}$ one), the flux received at wavelengths $> 100 $ $\mu$m  traces the FHSC outer boundary, and the accretion luminosity can escape. This result clearly shows the importance of properly resolving the accretion shock on the FHSC with a full RHD solver and can explain the differences found in the past literature. It also has some important implications for observability since an FHSC should then be observable as a point source at mid- to far-infrared wavelengths if the inclination is favorable.
  
 Last but not least, there is a clear FHSC lifetime dependance on the initial magnetization of the cloud. Whereas the FHSC has a lifetime $> 4$ kyr in the MU200 model, it is shorter in the other models as magnetization increases, i.e.,  $\sim 3$ kyr for MU10 and  $\sim 1.2$ kyr for MU2. The evolution is faster as magnetization increases, since the accretion rate increases with stronger magnetic braking that transports angular momentum outward \citep{Commercon_et_al_2010,Joos_et_al_2012}. As already mentioned, \cite{Tomida_2010b} reached a similar conclusion about the FHSC lifetime using RHD models of different initial mass rotating dense cores.

\subsection{$L_\mathrm{bol}$, $T_\mathrm{bol}$, and $L_\mathrm{bol}/L_\mathrm{smm}$}

\begin{figure*}[t]
  \centering
  \includegraphics[scale=0.8]{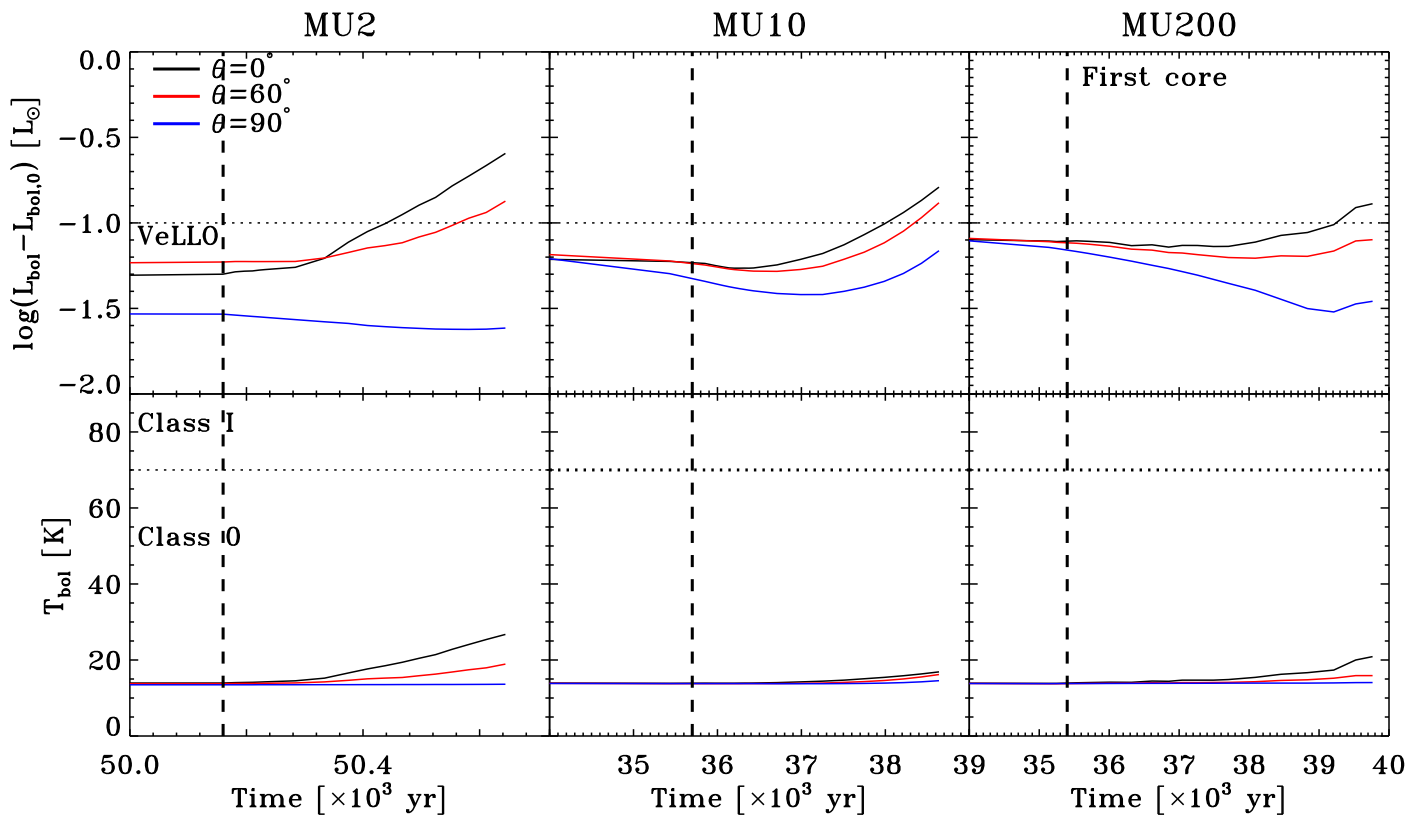}
\caption{Bolometric luminosity (top) and bolometric temperature (bottom) as a function of time for the three models: MU2 (left), MU10 (middle), and MU200 (right). Bolometric quantities are computed from models with three different inclination angles: $\theta=0^{\circ}$ (black), $\theta=60^{\circ}$ (red), and $\theta=90^{\circ}$ (blue).}
\label{ltbol_time}
\end{figure*}

Figure \ref{ltbol_time} shows the evolution of the internal luminosity, i.e., the difference between the  bolometric luminosity at time $t$ and the initial bolometric luminosity (top panels) and the bolometric temperature (bottom panels) as a function of time for the three models and three inclination angles: $\theta=0^{\circ}$, $60^{\circ}$, and $90^{\circ}$. Time $t_0$ is indicated by the vertical dashed line. 
The dotted line in the bolometric luminosity plots indicates the VeLLO boundary \citep{DiFrancesco_et_al_2007}, $L_\mathrm{int}<0.1$ L$_\odot$. The bolometric luminosity and temperature globally increase after the formation of  the FHSC for inclination angles $\theta<60^{\circ}$. Only the edge-on view gives rise to decreasing bolometric luminosity because of the disk or pseudo-disk  structure. More interestingly, the luminosity in the $\theta<60^{\circ}$ cases increases relatively slowly and becomes greater than $0.1$ L$_\odot$ only at late times, long after the FHSC formation.   FHSCs will thus appear as VeLLOs for most or all of their lifetime.

In the bolometric temperature plots, the horizontal dotted line shows the Class 0/Class I boundary of \cite{Chen_et_al_1995}, i.e., $T_\mathrm{bol}=70$ K. In terms of $T_\mathrm{bol}$, FHSCs are indistinguishable from starless cores and remain at the lower end of the Class 0 $T_\mathrm{bol}$ range. The same behavior is obtained for the ratio $ L_\mathrm{bol}/L_\mathrm{smm}$. In summary, these classical definitions cannot help in distinguishing between an FHSC and a more evolved Class 0 source, or even between an FHSC and a starless core. 


 \subsection{FHSC signature\label{new_FHSC}}
\begin{figure*}[]
  \centering
  \includegraphics[scale=0.9]{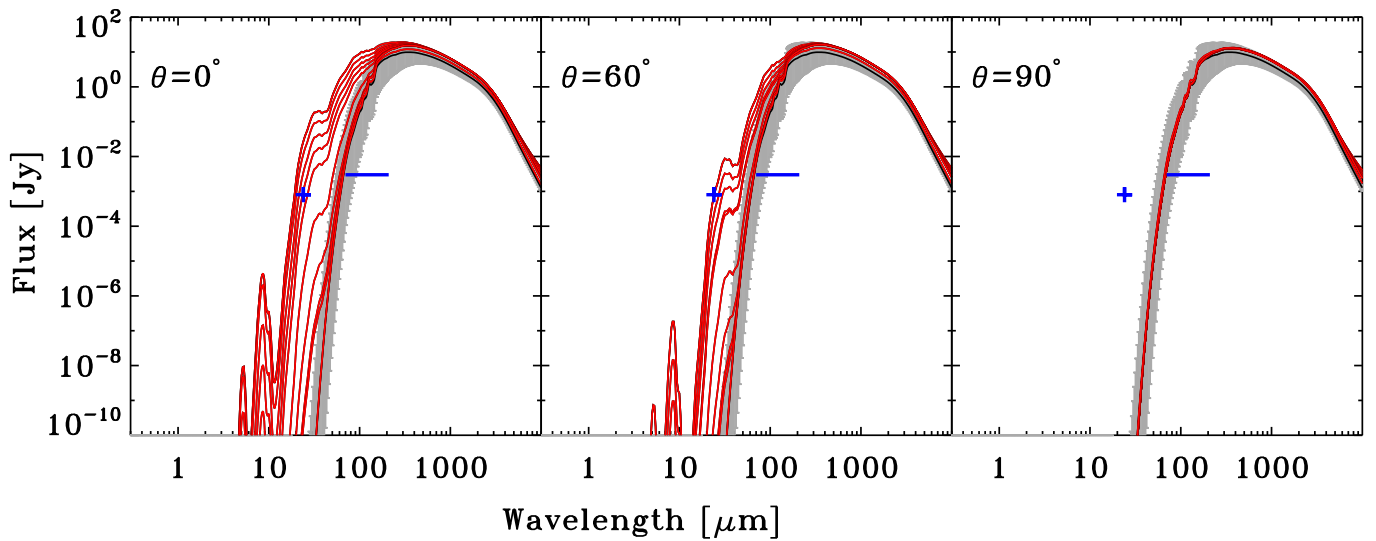}
\caption{Spectral energy distribution evolution as a function of time and inclination for the MU2 model. The black line represents the SED when the FHSC has not yet been formed, whereas the red lines indicate SEDs after FHSC formation. The gray area around the initial SED reflects the uncertainties in the initial temperature of the prestellar core (9 K$<T_0<$13 K). The blue cross represents the sensitivity at 24 $\mu$m ($0.83$ mJy at $3\sigma/24$ s) for the {\it{Spitzer }}\rm c2d Legacy program \citep{Evans_et_al_2003} and the horizontal blue line the sensitivity of the {\it{Herschel }}\rm PACS instrument \citep[$\sim 4.4$ mJy in point source mode, $5\sigma/1$h; ][]{Poglitsch_et_al_2010}.}
\label{SED_evol}
\end{figure*}

 \begin{figure*}[t]
  \centering
  \includegraphics[scale=0.8]{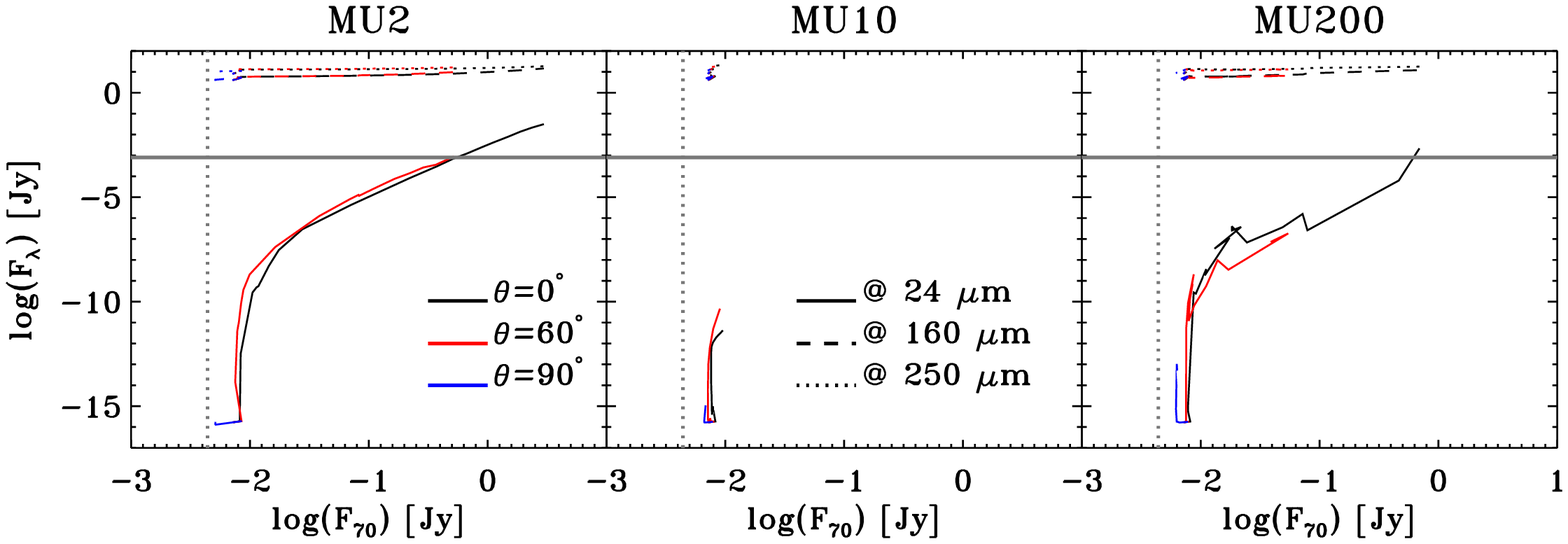}

  \includegraphics[scale=0.8]{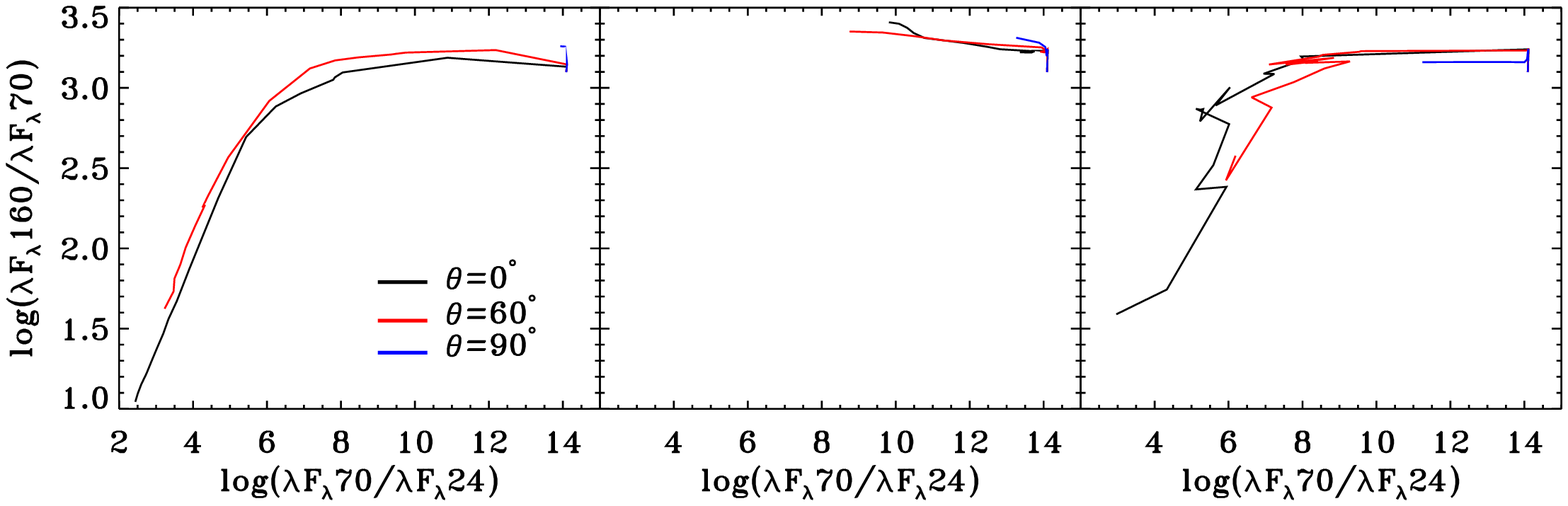}
\caption{{\it{Top: }}\rm Evolutionary diagram of the flux at at 24 $\mu$m (solid line), 160 $\mu$m (dotted line), and 250 $\mu$m (dashed line) as a function of the flux at 70 $\mu$m  for the three models and for the the same three inclination angles as previously. The horizontal gray line indicates the aforementioned  24 $\mu$m sensitivity limit of {\it{Spitzer}} and the vertical dotted line the 70 $\mu$m sensitivity limit of the {\it{Herschel }}\rm PACS instrument. {\it{ Bottom: }}\rm Color-color plot representing log($\lambda\mathrm{F}_\lambda160/\lambda\mathrm{
F}_\lambda70$) as a function of log($\lambda\mathrm{F}_\lambda70/\lambda\mathrm{F}_\lambda24$)  for the three models with the same three inclination angles.}
\label{L70}
\end{figure*}

\begin{figure*}[]
  \centering
  \includegraphics[scale=0.9]{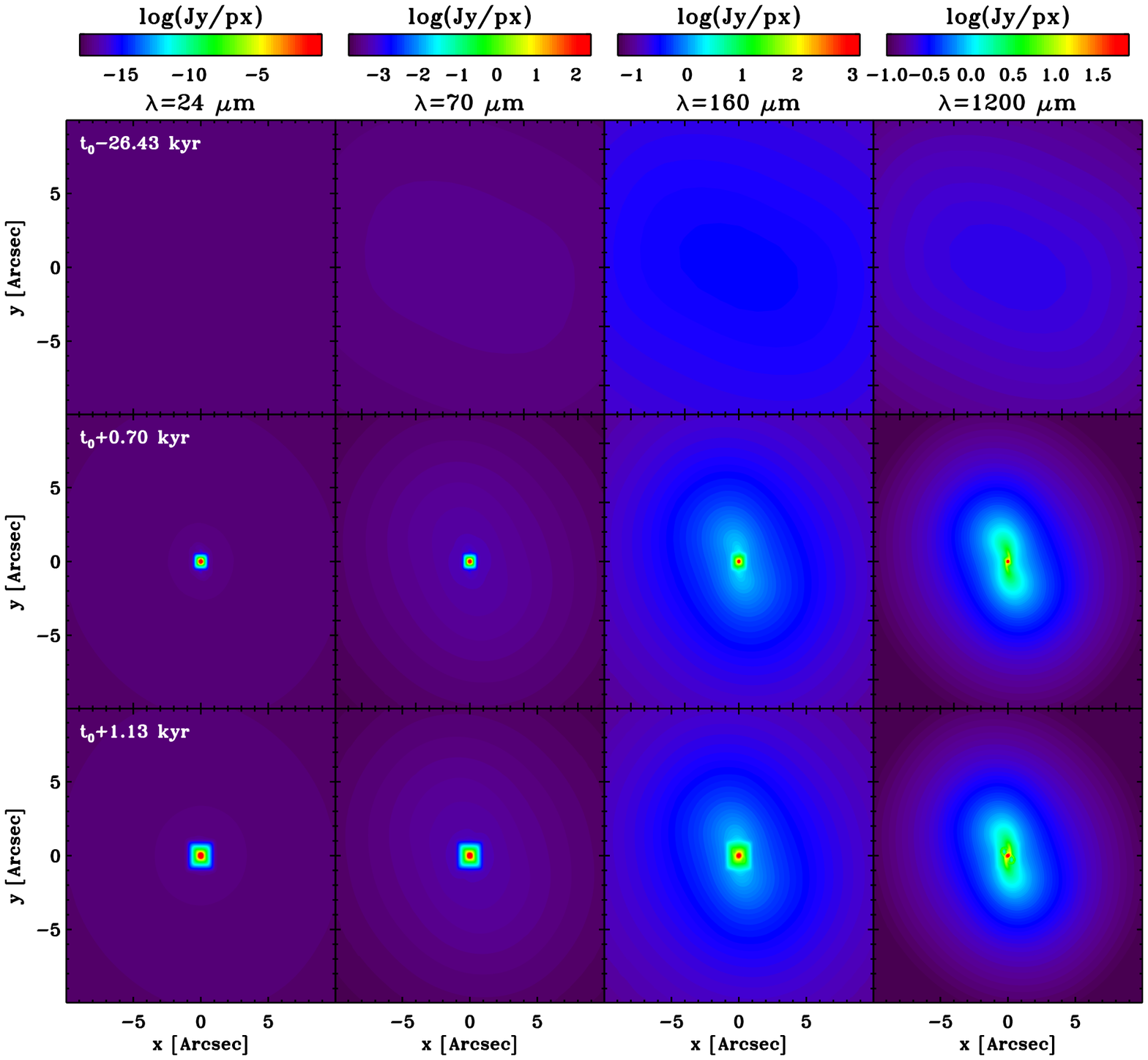}
\caption{Dust continuum emission maps for the MU2 models at four wavelengths, from left to right: 24 $\mu$m, 70 $\mu$m, 160 $\mu$m, and 1200 $\mu$m. Three different times are shown that trace the evolutionary stage of the source: $t_0-26.43$ kyr (top row, i.e., starless core), $t_0+0.7$ kyr (middle row, i.e., faint FHSC) and $t_0+1.13$ kyr (bottom row, i.e., end of the FHSC lifetime). The fluxes are given in Jy/px  (1 px = $0.08''\times0.08''$) for objects at a distance of 150 pc and observed pole-on. }
\label{mu2_images}
\end{figure*}
 
 We have shown in the previous section that classical observational signatures are not reliable evolutionary indicators during the early collapse stage. Thanks to the increasing spectral coverage and sensitivity of new instruments, however, SEDs are increasingly more  accurate, and new indicators can be defined to distinguish between a collapsing core with an FHSC and a prestellar (starless) core.
 
 Figure \ref{SED_evol}  shows the SED evolution as a function of time and inclination for the MU2 model at stages before (black lines) and after (red lines) FHSC formation. The blue cross represents the sensitivity at 24 $\mu$m ($0.83$ mJy at $3\sigma/24$ s) for the c2d Legacy program \citep{Evans_et_al_2003} and the horizontal blue line {represents the inflight sensitivity of the {\it{Herschel }} PACS instrument at 70 $\mu$m, derived from science observations \citep[$\sim 4.4$ mJy in point source mode, $5\sigma/1$h, ][]{Poglitsch_et_al_2010}. Shortly after FHSC formation, the SED does not  significantly depart from the initial one. After the FHSC begins to build up its mass, however, the accretion shock becomes stronger and departures from the initial SED are much clearer. To take into account the uncertainties on the initial temperature of the prestellar cores, we show SEDs of prestellar cores with initial temperatures ranging from 9 K to 13 K (about 20 \% upper limit uncertainty; Launhardt et al., {\it{in prep}}\rm). The SEDs with accreting FHSC are clearly distinct from the initial uncertainties on the temperature for inclination $\theta<60^{\circ}$ although the SED peak itself does not show evidence of an FHSC. According to the sensitivity of the {\it{Herschel }}\rm PACS instrument, FHSCs can be identified with sufficiently deep observations. More interestingly, the sensitivity of {\it{Spitzer }}\rm at 24 $\mu$m could also be sufficient  for FHSC detection at late stages of their lifetime.  The wavelength range over which the flux has increased the most after the FHSC formation is 20 $\mu$m $<\lambda<$ 100 $\mu$m (at least a factor 10 at 100 $\mu$m for $\theta<60^{\circ}$), whereas $L_\mathrm{smm}$ increases only by  a factor less than 2. Similarly, the ratio $L_\mathrm{bol}/L_\mathrm{smm}$ increases at maximum by only a factor $\sim 2$ after FHSC formation. 

The estimate of $L_\mathrm{bol}$ from observations strongly depends on the number of points in the SED, i.e., the spectral coverage. Instead of the physical stage indicator $L_\mathrm{bol}/L_\mathrm{smm}$, we chose here to express a related luminosity ratio using a particular wavelength. In this example, we chose the luminosity at 70 $\mu$m to show how it is possible with the {\it{Spitzer}}\rm/MIPS or {\it{Herschel}}\rm/PACS instruments to identify FHSC candidates in comparison to starless cores. We note that a similar analysis can be performed using the 100 $\mu$m, depending on the available observations. The flux at this wavelength is well constrained, and we just showed that $L_\mathrm{smm}$ increases only by a factor 2 after the formation of the FHSC. Finally, we note that similar flux increases in the wavelength range 20 $\mu$m $<\lambda<$ 100 $\mu$m are also observed in 10 M$_\odot$ dense core collapse calculations since, even if the envelope is thicker, the accretion shock on the FHSC is also stronger \citep[e.g., for massive dense core models, see ][]{Commercon_et_al_2011c}. 
 
 Figure \ref{L70} (top) shows the time evolution  of the flux at 24 $\mu$m (solid line), 160 $\mu$m (dotted line), and 1200 $\mu$m (dashed line) as a function of the flux at 70 $\mu$m and of the inclination angle for the three models.  The horizontal gray line indicates the aforementioned  24 $\mu$m sensitivity limit of {\it{Spitzer }}\rm and the vertical dotted line the 70 $\mu$m sensitivity limit of the {\it{Herschel }}\rm PACS instrument. 
For $\theta<60^{\circ}$, the flux  $F_\mathrm{70}$ increases by about two orders of magnitude after FHSC formation for the MU2  and MU200 models, whereas the MU10 model does not show any increase, as expected, since its SED does not show any evolution. Note that the flux received at 70 $\mu$m is always above the sensitivity limits of the current instruments. The most dramatic increase is found in the flux at 24 $\mu$m, which is almost zero at FHSC formation and then increases to detectable values at late times. Similar to $L_\mathrm{smm}$, the fluxes at 160 $\mu$m and 1200 $\mu$m remain almost constant throughout the FHSC lifetime. The fluxes at 24 $\mu$m and 70 $\mu$m can thus help in identifying FHSC candidates with respect to starless cores if the physical conditions are appropriate (either a strong or a very weak magnetic field), and if the objects are observed with  a favorable inclination $<60^{\circ}$. The evolutionary tracks of the FHSCs in the color-color plane of log($\lambda\mathrm{F}_\lambda70/\lambda\mathrm{F}_\lambda24$)-log($\lambda\mathrm{F}_\lambda160/\lambda\mathrm{F}_\lambda70$) are shown in Fig. \ref{L70} (bottom) for the three models at the three same inclination angles as discussed previously. There is a clear evolutionary track followed by the accreting FHSC that can be directly compared to observations.

\section{Discussion}

\subsection{Comparison to previous work}

Spectral energy distributions of FHSCs embedded in their parent cores have already been studied for instance by \cite{Boss_Yorke_1995}, \cite{Masunaga_et_al_1998}, \cite{Omukai_2007}, \cite{Tomida_2010b}, and \cite{Saigo_Tomisaka_2011}.  While they all found that the radiation emitted by the FHSC is essentially completely reprocessed by the dust in the surrounding core
to 50 - 200 $\mu$m with no observable emission below 30 $\mu$m, we found that there is indeed observable emission at shorter far-infrared wavelengths down to 20 $\mu$m. These differences may be explained by the different physical models and post-processing of each study. While we directly post-processed 3D AMR RMHD calculations, \cite{Boss_Yorke_1995} used a spherical grid and integrated the RHD equations with a low numerical resolution. Their FHSC temperature is also relatively low ($<200$ K). Their results are therefore similar to ours shortly after FHSC formation. In addition, as mentioned previously, \cite{Masunaga_et_al_1998} and  \cite{Omukai_2007} used spherical models and their SEDs are then similar to ours with $\theta=90^{\circ}$. Next, \cite{Saigo_Tomisaka_2011} post-processed 3D high-resolution collapse calculations with plane-parallel approximation but neglected magnetic fields and used a barotropic EOS. They  observed an increase of the total energy in the SED but their SEDs were calculated over a very small area of 40 AU $\times$ 40 AU and therefore excluded the large-scale emission dominant at wavelengths $> 100$ $\mu$m. Last, \cite{Tomida_2010b} computed SEDs from 3D RHD calculations but their cores do not fragment (slow rotation). In addition, they used Bonnor-Ebert initial density profiles in which accretion rates are lower and therefore FHSCs are fainter.

\subsection{What can SEDs tell about physical processes and evolutionary stages?}

Figures  \ref{sed} and \ref{SED_evol} show that the evolution of an FHSC throughout its lifetime can be reflected in observable changes of the SED, but only under certain conditions. Only the intermediate MU10 model does not show any evolutionary sequence because of the thick envelope structure consisting of a system disk + pseudo-disk + outflow. The MU2 and MU200 models show an evolutionary sequence very different than that of the MU10 model, but very similar to each other. The MU2 model is strongly magnetized and exhibits a pseudo-disk + outflow system, whereas the MU200 model fragments and exhibits a rotationally supported disk.  An FHSC with a clear  evolutionary sequence can then be observed in simple geometrical configurations and with inclination angle $\theta<60^{\circ}$. Nevertheless, it is impossible to distinguish between a magnetized and a non-magnetized scenario from a strict SED point of view. High angular resolution images of, at least, dust emission are required to distinguish between these scenarios. This kind of diagnostic can only be provided by interferometry at longer wavelengths (see Paper II).

We have shown than it can be relatively straightforward to distinguish a starless core from an FHSC thanks to flux received at mid- to far-infrared wavelengths, i.e., at 24 $\mu$m and 70 $\mu$m, but it remains hard to distinguish between an FHSC and a young stellar object (YSO) with a disk \citep{Pineda_et_al_2011}, i.e., an accreting second hydrostatic core (SHSC). This difficulty is due to the thick envelope in which the FHSC and SHSC are each embedded throughout the Class 0 phase that reprocesses  short-wavelength radiation to longer wavelengths. Here, models clearly fail to predict any detectable differences. Using only dust continuum emission is not the appropriate way to differentiate the two core stages, since almost all radiation emanating from both the FHSC and the SHSC will be reprocessed by the thick envelope. Another approach could consist of focusing on the outflow properties (collimation, velocity, momentum, etc.) and on the chemical properties of the sources, such as gas phase and ice molecules abundances, which reflect the local thermal history \citep[e.g., ][]{Kim_et_al_2011}. We will study these properties in upcoming papers.

\subsection{Example of methodology to target first core}

We showed in Sec. \ref{new_FHSC} that the fluxes at 24 $\mu$m and 70 $\mu$m can be good indicators of FHSCs in comparison with starless cores. We also did not find any significant emission below 10 $\mu$m. We infer that the combination of a point source detection at 70 $\mu$m or even at 24 $\mu$m with no detection at wavelength $<10$ $\mu$m can be a good indicator of FHSCs from a strict SED point of view. Under the current telescope sensitivity limits, a detection at 70 $\mu$m alone is most likely, since the emission at 24 $\mu$m is faint, and reaches the {\it{Spitzer }} sensitivity limit only in the MU2 model in the second half of the FHSC lifetime. Figure \ref{mu2_images} shows dust continuum emission maps for the MU2 model at three different evolutionary stages and for four wavelengths (24 $\mu$m, 70 $\mu$m, 160 $\mu$m, and 1200 $\mu$m). It clearly shows that the emission at 70 $\mu$m and 24 $\mu$m should be point-like after the FHSC formation. Note that at all evolutionary stages, a flux at 70 $\mu$m can be detected but its emission would not look like a point source for a starless core. It will instead be much more extended (i.e., the size of the initial dense core) because of the ISRF reprocessing. 
Thanks to the recent {\it{Herschel }}\rm and {\it{Spitzer }}\rm observations, observers should be able to combine data from both telescopes and make comparisons with our predicted color-color plot in Fig. \ref{L70}. For instance,  \cite{DiFransesco_et_al_2010} already used this procedure and combined maps of {\it{Herschel }}\rm data with {\it{Spitzer }}\rm ones to distinguish between prestellar and protostellar cores. Once FHSC candidates are identified from these flux examinations, the next step will be to target them with high-resolution interferometry such as ALMA (see Paper II) to constrain the physical properties (e.g., strength of the magnetic field).

As mentioned earlier, a complementary approach would be to use chemical analysis to separate the FHSC and the SHSC stages. At temperatures above $\sim 1000$ K, grain evaporation starts and refractory chemical species such as C and Si are released into the gas phase, considerably changing the chemical network and the opacities \citep[e.g., ][]{Lenzuni_et_al_1995}. Deuterium-related chemistry could also trace an SHSC because deuterium-bearing species should be under-abundant in high-temperature regions \citep{Albertsson_et_al_2011}.

Another more direct and more promising way to stage differentiation would be to find evidence of a low-velocity outflow without its high-velocity counterparts. Recent theoretical studies have shown that a low-velocity outflow can be launched during the FHSC stage \citep[e.g., ][]{Hennebelle_Fromang_2008, Commercon_et_al_2010,Tomida_2010} without the high-velocity outflow that is always present at the SHSC stage \citep{Machida_et_al_2008}. Not only may the morphology and energetics of the outflow help in identifying the cores, but also the chemical composition of the gas in the outflow. Since temperatures are different during the FHSC and SHSC stages, we may expect that the low-velocity outflow would show evidence of  low-temperature gas-grain chemistry and that the high-velocity outflow would show evidence of high-temperature chemistry. In addition,  the low-velocity outflow is launched from regions with temperatures $<1000$ K where grains are still not fully evaporated. Grains should then be present in the FHSC outflow and since the velocity is relatively low (a few km s$^{-1}$), they are not expected to be destroyed in the shock at the envelope interface. Therefore, we may expect FHSC outflows to be seen in thermal dust emission if they are dense enough. It is worth mentioning that some dust continuum observations may be also contaminated by CO emission from outflow gas, making them easier to see \citep[e.g., ][]{Drabek_et_al_2012}. 

Recent work by \cite{Price_2012} reported the formation of a collimated jet driven from the FHSC, supporting recent observations of an FHSC candidate in Per-Bolo 58 \citep{Dunham_et_al_2011}. The timescale needed to produce such a collimated jet ($\sim 2$ kyr), however, is longer than the expected lifetime of FHSCs in a magnetized collapsing dense core. In addition, \cite{Price_2012} used sink particles (scales smaller than 5 AU are not described) and a barotropic approximation with an adiabatic index of $\gamma=7/5$, for which FHSC lifetimes are even shorter than those we predict in this study with $\gamma=5/3$ (the FHSC contracts adiabatically faster with $\gamma=7/5$). It is therefore very likely that FHSCs reported in \cite{Price_2012} should have already undergone second collapse and formed an SHSC within $\sim 2$ kyr after their formation.

\subsection{Limits of the model}

In this study, we used the gray radiative transfer approximation for the RMHD calculations which may not be the most appropriate method at later stages of the evolutions. As mentioned earlier, however, \cite{Vaytet_et_al_2011} showed that multi-frequency collapse calculations were well-reproduced by the gray FLD calculations of \cite{Commercon_et_al_2011b}, at least for the first collapse and FHSC stages. In addition, we used a constant  ratio of specific heats $\gamma$, set to $5/3$, although it is known that $\gamma$ should change to a value of $7/5$ as temperature increases, and H$_2$ rotational degrees of freedom become available \citep[e.g., ][]{Machida_et_al_2008}. With $\gamma=7/5$, FHSCs compress and cool more efficiently and thus reach   the second collapse more quickly \citep[e.g., ][]{Commercon_et_al_2010}, which reduces the FHSC lifetime.

For the {\ttfamily{RADMC-3D}} post-processing, we used one specific opacity model of \cite{Semenov_et_al_2003A&A}  for all temperatures and densities. This choice may not be appropriate since grain mantle compositions change as temperature increases. Nevertheless, the changes in opacity remain relatively minor at temperatures $<1000$ K and we do not expect them to have a significant impact on our results. 
Finally, we have neglected  scattering although it has been found to be important for wavelengths $<10$ $\mu$m \citep{Young_Evans_2005,Dunham_et_al_2010}. Since we did not find any emission from FHSCs at these wavelengths and we do not consider this wavelength range significant for our conclusions, scattering effects are not relevant for our purposes.

\section{Summary and perspectives}

We used state-of-the-art radiation-magneto-hydrodynamic models of the first collapse and first hydrostatic core formation in the context of low-mass star formation to predict spectral energy distributions and related observational indicators. We investigated the collapse and the fragmentation of  a 1 M$_\odot$ dense core with three different magnetization levels up to the start of the second collapse. We showed that the FHSC lifetime strongly depends on the magnetization of the collapsing core and that a strong magnetic field leads to a shorter lifetime because  magnetic braking raises the accretion rate. We post-processed the RMHD calculations with a 3D radiative transfer code and produced synthetic spectral energy distributions. We showed that FHSCs can be identified thanks to the rapid increase in the flux received between 20 $\mu$m and 100 $\mu$m for inclination angle $\theta<60^{\circ}$ under different physical conditions (e.g., high or weak magnetization). We also found that FHSCs are globally consistent with VeLLOs. The classical stage evolutionary indicators $T_\mathrm{bol}$, $L_\mathrm{bol}$, and $L_\mathrm{bol}/L_\mathrm{smm}$ are not reliable for identifying FHSCs since they do not show any significant evolution through the FHSC lifetime. We found that a point source detection at wavelengths shorter than 200 $\mu$m is a good indicator of FHSC in combination with the absence of detection at wavelengths $< 10$ $\mu$m . We currently cannot, however, provide any diagnostics to distinguish an FHSC from a low-luminosity SHSC.

Recently, {\it{Herschel }}\rm has mapped hundreds of dense cores through, e.g.,  the EPOS and Gould Belt Survey key observation programs \citep[e.g., ][]{Henning_et_al_2010,Andre_et_al_2010,Bontemps_et_al_2010} with PACS data at 70 $\mu$m. The community has also gathered many {\it{Spitzer }}\rm observations at 24 $\mu$m and 70 $\mu$m. Combining the observations for common sources of both datasets could identify FHSC candidates. Cataloging these candidates will provide the basis required for observation proposals aimed at high-resolution interferometers such as the {\it{Submillimeter Array }}\rm (SMA) and ALMA, which will be indispensable to accurately measure the physical properties of the collapsing cores (see below). Future progress is also expected from the next generation of space telescopes such as SPICA with the SAFARI instrument \citep[with 1-sigma rms sensitivities of $\sim 2$  $\mu$Jy at 24 $\mu$m and $\sim 50$  $\mu$Jy at 70 $\mu$m; ][]{Swinyard_2009}, which could then test the evolutionary tracks we predicted.

SED information alone cannot help in distinguishing the important physical mechanisms at work at early Class 0 stages (large disk formation and fragmentation, outflow launching, etc.). Only interferometric observations would be capable of such a feat, and forthcoming ALMA observations will give a clearer, if not definitive, answer to the fragmentation properties of Class 0 sources. ALMA dust continuum emission synthetic maps  will be presented in Paper II.  Predictions other than dust continuum emission only are also needed to target FHSC candidates more efficiently. Chemistry post-processing and molecular line emission calculations are therefore a natural next step for FHSC observation predictions. 
Future investigations will also aim to follow the second collapse with the appropriate physics (mainly H$_2$ dissociation).

\begin{acknowledgements}
We thank the anonymous referee for her/his comments and corrections that improved the quality of the manuscript. The calculations were performed at CEA on the DAPHPC cluster. The research of B.C. is supported by the postdoctoral fellowships from Max-Planck-Institut f\"{u}r Astronomie and from the CNES. BC also acknowledges support from the French ANR Retour Postdoc program. BC thanks N. Vaytet, F. Levrier and A. Stutz for useful comments and discussions.
\end{acknowledgements}
\bibliographystyle{aa}
\bibliography{biblio1}

\begin{thebibliography}{61}
\expandafter\ifx\csname natexlab\endcsname\relax\def\natexlab#1{#1}\fi

\bibitem[{{Albertsson} {et~al.}(2011){Albertsson}, {Semenov}, \&
  {Henning}}]{Albertsson_et_al_2011}
{Albertsson}, T., {Semenov}, D.~A., \& {Henning}, T. 2011, ArXiv e-prints

\bibitem[{{Andr{\'e}} {et~al.}(2010){Andr{\'e}}, {Men'shchikov}, {Bontemps},
  {K{\"o}nyves}, {Motte}, {Schneider}, {Didelon}, {Minier}, {Saraceno},
  {Ward-Thompson}, {di Francesco}, {White}, {Molinari}, {Testi}, {Abergel},
  {Griffin}, {Henning}, {Royer}, {Mer{\'{\i}}n}, {Vavrek}, {Attard},
  {Arzoumanian}, {Wilson}, {Ade}, {Aussel}, {Baluteau}, {Benedettini},
  {Bernard}, {Blommaert}, {Cambr{\'e}sy}, {Cox}, {di Giorgio}, {Hargrave},
  {Hennemann}, {Huang}, {Kirk}, {Krause}, {Launhardt}, {Leeks}, {Le Pennec},
  {Li}, {Martin}, {Maury}, {Olofsson}, {Omont}, {Peretto}, {Pezzuto}, {Prusti},
  {Roussel}, {Russeil}, {Sauvage}, {Sibthorpe}, {Sicilia-Aguilar}, {Spinoglio},
  {Waelkens}, {Woodcraft}, \& {Zavagno}}]{Andre_et_al_2010}
{Andr{\'e}}, P., {Men'shchikov}, A., {Bontemps}, S., {et~al.} 2010, \aap, 518,
  L102+

\bibitem[{{Andr{\'e}} {et~al.}(1993){Andr{\'e}}, {Ward-Thompson}, \&
  {Barsony}}]{Andre_et_al_1993}
{Andr{\'e}}, P., {Ward-Thompson}, D., \& {Barsony}, M. 1993, \apj, 406, 122

\bibitem[{{Belloche} {et~al.}(2006){Belloche}, {Parise}, {van der Tak},
  {Schilke}, {Leurini}, {G{\"u}sten}, \& {Nyman}}]{Belloche_et_al_2006}
{Belloche}, A., {Parise}, B., {van der Tak}, F.~F.~S., {et~al.} 2006, \aap,
  454, L51

\bibitem[{{Bontemps} {et~al.}(2010){Bontemps}, {Andr{\'e}}, {K{\"o}nyves},
  {Men'shchikov}, {Schneider}, {Maury}, {Peretto}, {Arzoumanian}, {Attard},
  {Motte}, {Minier}, {Didelon}, {Saraceno}, {Abergel}, {Baluteau}, {Bernard},
  {Cambr{\'e}sy}, {Cox}, {di Francesco}, {di Giorgo}, {Griffin}, {Hargrave},
  {Huang}, {Kirk}, {Li}, {Martin}, {Mer{\'{\i}}n}, {Molinari}, {Olofsson},
  {Pezzuto}, {Prusti}, {Roussel}, {Russeil}, {Sauvage}, {Sibthorpe},
  {Spinoglio}, {Testi}, {Vavrek}, {Ward-Thompson}, {White}, {Wilson},
  {Woodcraft}, \& {Zavagno}}]{Bontemps_et_al_2010}
{Bontemps}, S., {Andr{\'e}}, P., {K{\"o}nyves}, V., {et~al.} 2010, \aap, 518,
  L85+

\bibitem[{{Boss} \& {Yorke}(1995)}]{Boss_Yorke_1995}
{Boss}, A.~P. \& {Yorke}, H.~W. 1995, \apjl, 439, L55

\bibitem[{{Chen} {et~al.}(1995){Chen}, {Myers}, {Ladd}, \&
  {Wood}}]{Chen_et_al_1995}
{Chen}, H., {Myers}, P.~C., {Ladd}, E.~F., \& {Wood}, D.~O.~S. 1995, \apj, 445,
  377

\bibitem[{{Chen} {et~al.}(2010){Chen}, {Arce}, {Zhang}, {Bourke}, {Launhardt},
  {Schmalzl}, \& {Henning}}]{Chen_et_al_2010}
{Chen}, X., {Arce}, H.~G., {Zhang}, Q., {et~al.} 2010, \apj, 715, 1344

\bibitem[{{Commer{\c c}on} {et~al.}(2011{\natexlab{a}}){Commer{\c c}on},
  {Audit}, {Chabrier}, \& {Chi{\`e}ze}}]{Commercon_et_al_2011b}
{Commer{\c c}on}, B., {Audit}, E., {Chabrier}, G., \& {Chi{\`e}ze}, J.-P.
  2011{\natexlab{a}}, \aap, 530, A13+

\bibitem[{{Commer{\c c}on} {et~al.}(2008){Commer{\c c}on}, {Hennebelle},
  {Audit}, {Chabrier}, \& {Teyssier}}]{Commercon_et_al_2008}
{Commer{\c c}on}, B., {Hennebelle}, P., {Audit}, E., {Chabrier}, G., \&
  {Teyssier}, R. 2008, \aap, 482, 371

\bibitem[{{Commer{\c c}on} {et~al.}(2010){Commer{\c c}on}, {Hennebelle},
  {Audit}, {Chabrier}, \& {Teyssier}}]{Commercon_et_al_2010}
{Commer{\c c}on}, B., {Hennebelle}, P., {Audit}, E., {Chabrier}, G., \&
  {Teyssier}, R. 2010, \aap, 510, L3+

\bibitem[{{Commer{\c c}on} {et~al.}(2011{\natexlab{b}}){Commer{\c c}on},
  {Hennebelle}, \& {Henning}}]{Commercon_et_al_2011c}
{Commer{\c c}on}, B., {Hennebelle}, P., \& {Henning}, T. 2011{\natexlab{b}},
  \apjl, 742, L9

\bibitem[{{Commer{\c c}on} {et~al.}(2011{\natexlab{c}}){Commer{\c c}on},
  {Teyssier}, {Audit}, {Hennebelle}, \& {Chabrier}}]{Commercon_et_al_2011a}
{Commer{\c c}on}, B., {Teyssier}, R., {Audit}, E., {Hennebelle}, P., \&
  {Chabrier}, G. 2011{\natexlab{c}}, \aap, 529, A35+

\bibitem[{{di Francesco} {et~al.}(2007){di Francesco}, {Evans}, {Caselli},
  {Myers}, {Shirley}, {Aikawa}, \& {Tafalla}}]{DiFrancesco_et_al_2007}
{di Francesco}, J., {Evans}, II, N.~J., {Caselli}, P., {et~al.} 2007,
  Protostars and Planets V, 17

\bibitem[{{di Francesco} {et~al.}(2010){di Francesco}, {Sadavoy}, {Motte},
  {Schneider}, {Hennemann}, {Csengeri}, {Bontemps}, {Balog}, {Zavagno},
  {Andr{\'e}}, {Saraceno}, {Griffin}, {Men'shchikov}, {Abergel}, {Baluteau},
  {Bernard}, {Cox}, {Deharveng}, {Didelon}, {di Giorgio}, {Hargrave}, {Huang},
  {Kirk}, {Leeks}, {Li}, {Marston}, {Martin}, {Minier}, {Molinari}, {Olofsson},
  {Persi}, {Pezzuto}, {Russeil}, {Sauvage}, {Sibthorpe}, {Spinoglio}, {Testi},
  {Teyssier}, {Vavrek}, {Ward-Thompson}, {White}, {Wilson}, \&
  {Woodcraft}}]{DiFransesco_et_al_2010}
{di Francesco}, J., {Sadavoy}, S., {Motte}, F., {et~al.} 2010, \aap, 518, L91+

\bibitem[{{Drabek} {et~al.}(2012){Drabek}, {Hatchell}, {Friberg}, {Richer},
  {Graves}, {Buckle}, {Nutter}, {Johnstone}, \& {Di
  Francesco}}]{Drabek_et_al_2012}
{Drabek}, E., {Hatchell}, J., {Friberg}, P., {et~al.} 2012, ArXiv e-prints

\bibitem[{{Dunham} {et~al.}(2011){Dunham}, {Chen}, {Arce}, {Bourke}, {Schnee},
  \& {Enoch}}]{Dunham_et_al_2011}
{Dunham}, M.~M., {Chen}, X., {Arce}, H.~G., {et~al.} 2011, \apj, 742, 1

\bibitem[{{Dunham} {et~al.}(2010){Dunham}, {Evans}, {Terebey}, {Dullemond}, \&
  {Young}}]{Dunham_et_al_2010}
{Dunham}, M.~M., {Evans}, II, N.~J., {Terebey}, S., {Dullemond}, C.~P., \&
  {Young}, C.~H. 2010, \apj, 710, 470

\bibitem[{{Enoch} {et~al.}(2010){Enoch}, {Lee}, {Harvey}, {Dunham}, \&
  {Schnee}}]{Enoch_et_al_2010}
{Enoch}, M.~L., {Lee}, J.-E., {Harvey}, P., {Dunham}, M.~M., \& {Schnee}, S.
  2010, \apjl, 722, L33

\bibitem[{{Evans} {et~al.}(2003){Evans}, {Allen}, {Blake}, {Boogert}, {Bourke},
  {Harvey}, {Kessler}, {Koerner}, {Lee}, {Mundy}, {Myers}, {Padgett},
  {Pontoppidan}, {Sargent}, {Stapelfeldt}, {van Dishoeck}, {Young}, \&
  {Young}}]{Evans_et_al_2003}
{Evans}, II, N.~J., {Allen}, L.~E., {Blake}, G.~A., {et~al.} 2003, \pasp, 115,
  965

\bibitem[{{Fromang} {et~al.}(2006){Fromang}, {Hennebelle}, \&
  {Teyssier}}]{Fromang_et_al_2006}
{Fromang}, S., {Hennebelle}, P., \& {Teyssier}, R. 2006, \aap, 457, 371

\bibitem[{{Galli} {et~al.}(2002){Galli}, {Walmsley}, \& {Gon{\c
  c}alves}}]{Galli_et_al_2002}
{Galli}, D., {Walmsley}, M., \& {Gon{\c c}alves}, J. 2002, \aap, 394, 275

\bibitem[{{Hennebelle} \& {Ciardi}(2009)}]{Hennebelle_Ciardi_2009}
{Hennebelle}, P. \& {Ciardi}, A. 2009, \aap, 506, L29

\bibitem[{{Hennebelle} {et~al.}(2011){Hennebelle}, {Commer{\c c}on}, {Joos},
  {Klessen}, {Krumholz}, {Tan}, \& {Teyssier}}]{Hennebelle_et_al_2011}
{Hennebelle}, P., {Commer{\c c}on}, B., {Joos}, M., {et~al.} 2011, \aap, 528,
  A72+

\bibitem[{{Hennebelle} \& {Fromang}(2008)}]{Hennebelle_Fromang_2008}
{Hennebelle}, P. \& {Fromang}, S. 2008, \aap, 477, 9

\bibitem[{{Hennebelle} \& {Teyssier}(2008)}]{Hennebelle_Teyssier_2008}
{Hennebelle}, P. \& {Teyssier}, R. 2008, \aap, 477, 25

\bibitem[{{Henning} {et~al.}(2010){Henning}, {Linz}, {Krause}, {Ragan},
  {Beuther}, {Launhardt}, {Nielbock}, \& {Vasyunina}}]{Henning_et_al_2010}
{Henning}, T., {Linz}, H., {Krause}, O., {et~al.} 2010, \aap, 518, L95+

\bibitem[{{Johnstone} {et~al.}(2000){Johnstone}, {Wilson}, {Moriarty-Schieven},
  {Joncas}, {Smith}, {Gregersen}, \& {Fich}}]{Johnstone_et_al_2000}
{Johnstone}, D., {Wilson}, C.~D., {Moriarty-Schieven}, G., {et~al.} 2000, \apj,
  545, 327

\bibitem[{{Joos} {et~al.}(2012){Joos}, {Hennebelle}, \&
  {Ciardi}}]{Joos_et_al_2012}
{Joos}, M., {Hennebelle}, P., \& {Ciardi}, A. 2012, ArXiv e-prints

\bibitem[{{Kim} {et~al.}(2011){Kim}, {Evans}, {Dunham}, {Chen}, {Lee},
  {Bourke}, {Huard}, {Shirley}, \& {De Vries}}]{Kim_et_al_2011}
{Kim}, H.~J., {Evans}, II, N.~J., {Dunham}, M.~M., {et~al.} 2011, \apj, 729, 84

\bibitem[{{Lada}(1987)}]{Lada_1987}
{Lada}, C.~J. 1987, in IAU Symposium, Vol. 115, Star Forming Regions, ed.
  {M.~Peimbert \& J.~Jugaku}, 1--17

\bibitem[{{Larson}(1969)}]{Larson_1969}
{Larson}, R.~B. 1969, \mnras, 145, 271

\bibitem[{{Launhardt} {et~al.}(2010){Launhardt}, {Nutter}, {Ward-Thompson},
  {Bourke}, {Henning}, {Khanzadyan}, {Schmalzl}, {Wolf}, \&
  {Zylka}}]{Launhardt_et_al_2010}
{Launhardt}, R., {Nutter}, D., {Ward-Thompson}, D., {et~al.} 2010, \apjs, 188,
  139

\bibitem[{{Lenzuni} {et~al.}(1995){Lenzuni}, {Gail}, \&
  {Henning}}]{Lenzuni_et_al_1995}
{Lenzuni}, P., {Gail}, H.-P., \& {Henning}, T. 1995, \apj, 447, 848

\bibitem[{{Machida} {et~al.}(2008){Machida}, {Inutsuka}, \&
  {Matsumoto}}]{Machida_et_al_2008}
{Machida}, M.~N., {Inutsuka}, S.-i., \& {Matsumoto}, T. 2008, \apj, 676, 1088

\bibitem[{{Masunaga} \& {Inutsuka}(2000)}]{Masunaga_Inutsuka_2000}
{Masunaga}, H. \& {Inutsuka}, S.-i. 2000, \apj, 531, 350

\bibitem[{{Masunaga} {et~al.}(1998){Masunaga}, {Miyama}, \&
  {Inutsuka}}]{Masunaga_et_al_1998}
{Masunaga}, H., {Miyama}, S.~M., \& {Inutsuka}, S.-I. 1998, \apj, 495, 346

\bibitem[{{Maury} {et~al.}(2010){Maury}, {Andr{\'e}}, {Hennebelle}, {Motte},
  {Stamatellos}, {Bate}, {Belloche}, {Duch{\^e}ne}, \&
  {Whitworth}}]{Maury_et_al_2010}
{Maury}, A.~J., {Andr{\'e}}, P., {Hennebelle}, P., {et~al.} 2010, \aap, 512,
  A40+

\bibitem[{{Mihalas} \& {Mihalas}(1984)}]{Mihalas_book}
{Mihalas}, D. \& {Mihalas}, B.~W. 1984, {Foundations of radiation
  hydrodynamics}, ed. D.~{Mihalas} \& B.~W. {Mihalas}

\bibitem[{{Miyoshi} \& {Kusano}(2005)}]{Miyoshi_Kusano_05}
{Miyoshi}, T. \& {Kusano}, K. 2005, Journal of Computational Physics, 208, 315

\bibitem[{{Motte} {et~al.}(1998){Motte}, {Andre}, \& {Neri}}]{Motte_et_al_1998}
{Motte}, F., {Andre}, P., \& {Neri}, R. 1998, \aap, 336, 150

\bibitem[{{Myers} \& {Ladd}(1993)}]{Myers_Ladd_1993}
{Myers}, P.~C. \& {Ladd}, E.~F. 1993, \apjl, 413, L47

\bibitem[{{Nutter} \& {Ward-Thompson}(2007)}]{Nutter_2007}
{Nutter}, D. \& {Ward-Thompson}, D. 2007, \mnras, 374, 1413

\bibitem[{{Omukai}(2007)}]{Omukai_2007}
{Omukai}, K. 2007, \pasj, 59, 589

\bibitem[{{Pineda} {et~al.}(2011){Pineda}, {Arce}, {Schnee}, {Goodman},
  {Bourke}, {Foster}, {Robitaille}, {Tanner}, {Kauffmann}, {Tafalla},
  {Caselli}, \& {Anglada}}]{Pineda_et_al_2011}
{Pineda}, J.~E., {Arce}, H.~G., {Schnee}, S., {et~al.} 2011, \apj, 743, 201

\bibitem[{{Poglitsch} {et~al.}(2010){Poglitsch}, {Waelkens}, {Geis},
  {Feuchtgruber}, {Vandenbussche}, {Rodriguez}, {Krause}, {Renotte}, {van
  Hoof}, {Saraceno}, {Cepa}, {Kerschbaum}, {Agn{\`e}se}, {Ali}, {Altieri},
  {Andreani}, {Augueres}, {Balog}, {Barl}, {Bauer}, {Belbachir}, {Benedettini},
  {Billot}, {Boulade}, {Bischof}, {Blommaert}, {Callut}, {Cara}, {Cerulli},
  {Cesarsky}, {Contursi}, {Creten}, {De Meester}, {Doublier}, {Doumayrou},
  {Duband}, {Exter}, {Genzel}, {Gillis}, {Gr{\"o}zinger}, {Henning},
  {Herreros}, {Huygen}, {Inguscio}, {Jakob}, {Jamar}, {Jean}, {de Jong},
  {Katterloher}, {Kiss}, {Klaas}, {Lemke}, {Lutz}, {Madden}, {Marquet},
  {Martignac}, {Mazy}, {Merken}, {Montfort}, {Morbidelli}, {M{\"u}ller},
  {Nielbock}, {Okumura}, {Orfei}, {Ottensamer}, {Pezzuto}, {Popesso},
  {Putzeys}, {Regibo}, {Reveret}, {Royer}, {Sauvage}, {Schreiber}, {Stegmaier},
  {Schmitt}, {Schubert}, {Sturm}, {Thiel}, {Tofani}, {Vavrek}, {Wetzstein},
  {Wieprecht}, \& {Wiezorrek}}]{Poglitsch_et_al_2010}
{Poglitsch}, A., {Waelkens}, C., {Geis}, N., {et~al.} 2010, \aap, 518, L2

\bibitem[{{Price} \& {Bate}(2009)}]{Price_Bate_2009}
{Price}, D.~J. \& {Bate}, M.~R. 2009, \mnras, 398, 33

\bibitem[{{Price} {et~al.}(2012){Price}, {Tricco}, \& {Bate}}]{Price_2012}
{Price}, D.~J., {Tricco}, T.~S., \& {Bate}, M.~R. 2012, \mnras, 423, L45

\bibitem[{{Saigo} \& {Tomisaka}(2011)}]{Saigo_Tomisaka_2011}
{Saigo}, K. \& {Tomisaka}, K. 2011, \apj, 728, 78

\bibitem[{{Semenov} {et~al.}(2003){Semenov}, {Henning}, {Helling}, {Ilgner}, \&
  {Sedlmayr}}]{Semenov_et_al_2003A&A}
{Semenov}, D., {Henning}, T., {Helling}, C., {Ilgner}, M., \& {Sedlmayr}, E.
  2003, \aap, 410, 611

\bibitem[{{Shetty} {et~al.}(2011){Shetty}, {Glover}, {Dullemond}, \&
  {Klessen}}]{Shetty_2011}
{Shetty}, R., {Glover}, S.~C., {Dullemond}, C.~P., \& {Klessen}, R.~S. 2011,
  MNRAS, 11

\bibitem[{{Swinyard} {et~al.}(2009){Swinyard}, {Nakagawa}, {Merken}, {Royer},
  {Souverijns}, {Vandenbussche}, {Waelkens}, {Davis}, {Di Francesco},
  {Halpern}, {Houde}, {Johnstone}, {Joncas}, {Naylor}, {Plume}, {Scott},
  {Abergel}, {Bensammar}, {Braine}, {Buat}, {Burgarella}, {Cais}, {Dole},
  {Duband}, {Elbaz}, {Gerin}, {Giard}, {Goicoechea}, {Joblin}, {Jones},
  {Kneib}, {Lagache}, {Madden}, {Pons}, {Pajot}, {Rambaud}, {Ravera},
  {Ristorcelli}, {Rodriguez}, {Vives}, {Zavagno}, {Geis}, {Krause}, {Lutz},
  {Poglitsch}, {Raab}, {Stegmaier}, {Sturm}, {Tuffs}, {Lee}, {Koo}, {Im},
  {Pak}, {Han}, {Park}, {Nam}, {Jin}, {Lee}, {Yuk}, {Lee}, {Aikawa}, {Arimoto},
  {Doi}, {Enya}, {Fukagawa}, {Furusho}, {Hasegawa}, {Hayashi}, {Honda
  Kanagawa}, {Ida}, {Imanishi}, {Masatoshi}, {Inutsuka}, {Izumiura}, {Kamaya},
  {Kaneda}, {Kasuga}, {Kataza}, {Kawabata}, {Kawada}, {Kawakita}, {Kii},
  {Koda}, {Kodama}, {Kokubo}, {Komatsu}, {Matsuhara}, {Matsumoto}, {Matsuura},
  {Miyata}, {Miyata}, {Nagata}, {Nagata}, {Nakajima}, {Naoto}, {Nishi}, {Noda},
  {Okamoto}, {Okamoto}, {Omukai}, {Onaka}, {Ootsubo}, {Ouchi}, {Saito}, {Sato},
  {Sako}, {Sekiguchi}, {Shibai}, {Sugita}, {Sugitani}, {Susa}, {Pyo}, {Tamura},
  {Ueda}, {Ueno}, {Wada}, {Watanabe}, {Yamada}, {Yamamura}, {Yoshida},
  {Yoshimi}, {Yui}, {Benedettini}, {Cerulli}, {Di Giorgio}, {Molinari},
  {Orfei}, {Pezzuto}, {Piazzo}, {Saraceno}, {Spinoglio}, {de Graauw}, {de
  Korte}, {Helmich}, {Hoevers}, {Huisman}, {Shipman}, {van der Tak}, {van der
  Werf}, {Wild}, {Acosta-Pulido}, {Cernicharo}, {Herreros}, {Martin-Pintado},
  {Najarro}, {Perez-Fourmon}, {Ramon Pardo}, {Gomez}, {Castro Rodriguez},
  {Ade}, {Barlow}, {Clements}, {Ferlet}, {Fraser}, {Griffin}, {Griffin},
  {Hargrave}, {Isaak}, {Ivison}, {Mansour}, {Laniesse}, {Mauskopf}, {Morozov},
  {Oliver}, {Orlando}, {Page}, {Popescu}, {Serjeant}, {Sudiwala}, {Rigopoulou},
  {Walker}, {White}, {Viti}, {Winter}, {Bock}, {Bradford}, {Harwit}, \&
  {Holmes}}]{Swinyard_2009}
{Swinyard}, B., {Nakagawa}, T., {Merken}, P., {et~al.} 2009, Experimental
  Astronomy, 23, 193

\bibitem[{{Teyssier}(2002)}]{Teyssier_2002}
{Teyssier}, R. 2002, \aap, 385, 337

\bibitem[{{Teyssier} {et~al.}(2006){Teyssier}, {Fromang}, \&
  {Dormy}}]{Teyssier_et_al_2006}
{Teyssier}, R., {Fromang}, S., \& {Dormy}, E. 2006, Journal of Computational
  Physics, 218, 44

\bibitem[{{Tomida} {et~al.}(2010{\natexlab{a}}){Tomida}, {Machida}, {Saigo},
  {Tomisaka}, \& {Matsumoto}}]{Tomida_2010b}
{Tomida}, K., {Machida}, M.~N., {Saigo}, K., {Tomisaka}, K., \& {Matsumoto}, T.
  2010{\natexlab{a}}, \apjl, 725, L239

\bibitem[{{Tomida} {et~al.}(2010{\natexlab{b}}){Tomida}, {Tomisaka},
  {Matsumoto}, {Ohsuga}, {Machida}, \& {Saigo}}]{Tomida_2010}
{Tomida}, K., {Tomisaka}, K., {Matsumoto}, T., {et~al.} 2010{\natexlab{b}},
  \apj, 714, L58

\bibitem[{{Tomisaka} \& {Tomida}(2011)}]{Tomisaka_2011}
{Tomisaka}, K. \& {Tomida}, K. 2011, \pasj, 63, 1151

\bibitem[{{Vaytet} {et~al.}(2012){Vaytet}, {Audit}, {Chabrier}, {Commercon}, \&
  {Masson}}]{Vaytet_et_al_2011}
{Vaytet}, N., {Audit}, E., {Chabrier}, G., {Commercon}, B., \& {Masson}, J.
  2012, ArXiv e-prints

\bibitem[{{Ward-Thompson} {et~al.}(1994){Ward-Thompson}, {Scott}, {Hills}, \&
  {Andre}}]{Ward-Thompson_et_al_1994}
{Ward-Thompson}, D., {Scott}, P.~F., {Hills}, R.~E., \& {Andre}, P. 1994,
  \mnras, 268, 276

\bibitem[{{Yamada} {et~al.}(2009){Yamada}, {Machida}, {Inutsuka}, \&
  {Tomisaka}}]{Yamada_2009}
{Yamada}, M., {Machida}, M.~N., {Inutsuka}, S.-i., \& {Tomisaka}, K. 2009,
  \apj, 703, 1141

\bibitem[{{Young} \& {Evans}(2005)}]{Young_Evans_2005}
{Young}, C.~H. \& {Evans}, II, N.~J. 2005, \apj, 627, 293

\end{thebibliography}
\begin{appendix}

\section{AMR versus cartesian grid for postprocessing\label{appendixa}}

In this appendix, we present SEDs computed with two different post-processing methods. The first is the one we used throughout this paper, i.e., the AMR grid of {\ttfamily{RAMSES }}\rm is directly loaded in {\ttfamily{RADMC-3D}}\rm. The second projects the AMR data of {\ttfamily{RAMSES }}\rm into a uniform cube of extent $\sim 3000$ AU x 3000 AU. The second method is much simpler for interfacing, but a lot of information is lost when smoothing the finer levels onto the uniform grid. In our calculations, since the finer levels correspond to the high-temperature and high-density regions, smoothing to a lower resolution may affect the SEDs. 

Figure \ref{mu10_sed_res} shows SEDs for the MU10 models computed on three different grids: the AMR grid for the black lines, a uniform $256^3$ cube for the red lines, and a uniform $512^3$ cube for the blue lines. Two different inclination angles are shown: $\theta=0^{\circ}$ (solid line) and $\theta=90^{\circ}$ (dashed line). The three SEDs obtained with $\theta=90^{\circ}$ are similar with no emission below 30 $\mu$m independent of the grid used for post-processing, but they are different for $\theta=0^{\circ}$. For the latter case, there is indeed emission between 3 $\mu$m and 30 $\mu$m and an increase of the flux around 100 $\mu$m. We clearly see that poor resolution may strongly affect the SEDs, which solely comes from the fact that high-density regions have been smoothed out by reprojection of the AMR grid onto a uniform one.

\begin{figure}[thb]
  \centering
  \includegraphics[scale=0.43]{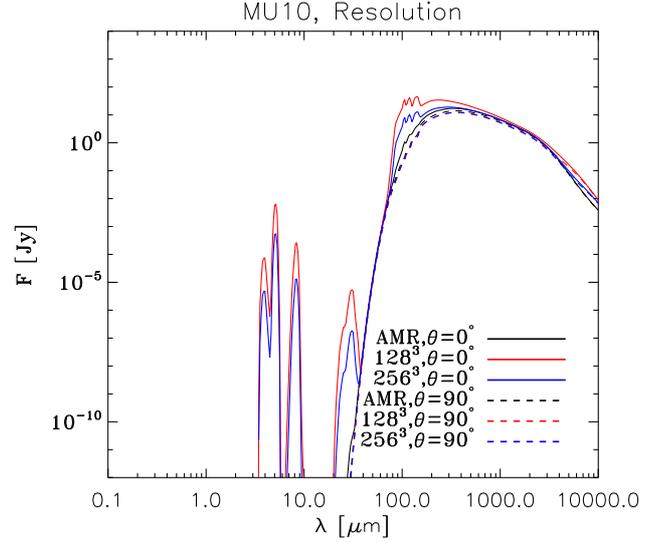}
\caption{Spectral energy distribution for the MU10 models computed on three different grids: the AMR grid for the black lines, a uniform $256^3$ cube for the red lines, and a uniform $512^3$ cube for the blue lines. Two different inclination angles are also shown: $\theta=0^{\circ}$ (solid line) and $\theta=90^{\circ}$ (dashed line, almost indistinguishable from one to another).}
\label{mu10_sed_res}
\end{figure}






\end{appendix}
\end{document}